\documentclass[a4paper,11pt]{article}
\usepackage{jheppub} % for details on the use of the package, please see the JINST-author-manual
\usepackage{lineno}
\usepackage{multirow}
\usepackage{orcidlink}
\input{belle2-symbols} 
\usepackage{overpic}
\usepackage{arydshln}
\newcommand{\notocsection}[1]{%
\refstepcounter{section}%
\section*{\thesection \quad #1}}%
\newcommand{\notocsubsection}[1]{%
\refstepcounter{subsection}%
\subsection*{\thesubsection \quad #1}}%

\arxivnumber{2404.12817} % if you have one

\title{\boldmath Determination of the CKM angle $\phi_{3}$ from a combination of Belle and Belle II results}

\collaboration{The Belle and Belle II Collaborations}
  \author{I.~Adachi\,\orcidlink{0000-0003-2287-0173},} % 2590
  \author{L.~Aggarwal\,\orcidlink{0000-0002-0909-7537},} % 10083
  \author{H.~Aihara\,\orcidlink{0000-0002-1907-5964},} % 2223
  \author{N.~Akopov\,\orcidlink{0000-0002-4425-2096},} % 9443
  \author{A.~Aloisio\,\orcidlink{0000-0002-3883-6693},} % 2194
  \author{S.~Al~Said\,\orcidlink{0000-0002-4895-3869},} % 6823
  \author{N.~Anh~Ky\,\orcidlink{0000-0003-0471-197X},} % 2218
  \author{D.~M.~Asner\,\orcidlink{0000-0002-1586-5790},} % 4684
  \author{H.~Atmacan\,\orcidlink{0000-0003-2435-501X},} % 2538
  \author{V.~Aushev\,\orcidlink{0000-0002-8588-5308},} % 2155
% ...
\emailAdd{coll-publications@belle2.org}
\abstract{We report a determination of the CKM angle $\phi_{3}$, also known as $\gamma$, from a combination of measurements using samples of up to 711~fb$^{-1}$ from the Belle experiment and up to 362~fb$^{-1}$ from the Belle II experiment. We combine results from analyses of $B^+\to DK^+, B^+\to D\pi^+$, and $B^+ \to D^{*}K^+$ decays, where $D$ is an admixture of $D^0$ and $\overline{D}{}^{0}$ mesons, in a likelihood fit to obtain $\phi_3 = (75.2\pm 7.6)^\circ$. We also briefly discuss the interpretation of this result.}

\begin{document}
\newcommand{\comment}[1]{\textcolor{red}{#1}}
\begin{flushright}
Belle II Preprint 2023-015 \\
KEK Preprint 2023-31
\end{flushright}
\maketitle
\flushbottom

%%%%%%%%%%%%%%%%%%%%%%%%%%%%%%%%%%%%%%%%%%%%%%%%%%%%%%%%%%%%%%%%%%%%%%%%%%%
\section{Introduction}
\label{sec:intro}
The Cabibbo-Kobayashi-Maskawa (CKM) matrix \cite{cabibbo,ckm} describes weak interactions of quarks and accommodates the only source of charge-parity symmetry ($\CP$) violation in the Standard Model (SM). The unitarity of the matrix can be represented as a triangle in the complex plane. One of its interior angles, $\phi_3$ (also known as $\gamma$), is defined as $\phi_3\equiv\arg{\left(-V_{ud}^{}V^{\ast}_{ub}/V_{cd}^{}V_{cb}^{\ast}\right)}$, where $V_{qq^{\prime}}$ are CKM matrix elements. The angle $\phi_3$ is of particular importance because it can be measured directly with negligible theoretical uncertainty by exploiting the interference between tree-level quark-transition amplitudes involving exchange of a single $W$-boson, $\bar{b}\to \bar{c}u\bar{s}$ and $\bar{b}\to \bar{u}c\bar{s}$~\cite{zupan}.\footnote{Throughout this paper, charge-conjugation is implied unless stated otherwise.} Hence, assuming only SM amplitudes in these processes, the measurement of $\phi_3$ provides an accurate reference to be compared against indirect determinations from global unitarity fits. The latter resulting from combinations of measurements of the sides and the other two angles of the unitarity triangle can be modified by non-SM particles via transitions involving more than one vector boson~\cite{blankeburas}. The comparison between direct and indirect determinations is thus a sensitive probe for the presence of non-SM particles in quark transitions. 
The current world average of direct measurements is $\left(66.4^{+2.8}_{-3.0}\right)^{\circ}$~\cite{pdg}, dominated by results from the LHCb collaboration~\cite{gammaLHCb}. 
From indirect determinations, the CKMfitter group obtains $\phi_3 = \left(66.29^{+0.72}_{-1.86}\right)^{\circ}$ \cite{ckmfitter}, while the UT$_{fit}$ Collaboration finds $\left(65.2\pm 1.5\right)^{\circ}$ \cite{UTfit:2022hsi}.
The difference in precision between direct and indirect determinations implies that an improvement in the former is important to better constrain non-SM contributions to $\CP$ violation. 

The angle $\phi_3$ is determined directly from the analysis of $B^+ \to Dh^+$ decays, where $D$ is an admixture of $D^0$ and $\overline{D}{}^{0}$ flavour eigenstates and $h$ is a kaon or pion.
The interference between the favoured $B^+ \to \overline{D}{}^{0}h^+$ decay amplitude mediated by a $b \to c$ transition, $\mathcal{A}_{\rm fav}$, and the suppressed $B^+ \to D^{0}h^+$ decay amplitude mediated by a $b \to u$ transition, $\mathcal{A}_{\rm sup}$, depends on $\phi_3$ and two hadronic parameters $r^X_B$ and $\delta^X_B$, which are given by the relation
\begin{equation}\label{eq:amplitude}
    \mathcal{A}_{\rm sup}/\mathcal{A}_{\rm fav} = r^X_B e^{i(\delta^X_B + \phi_3)},
\end{equation}
where $r_B$ and $\delta_B$ represent the decay-amplitude ratio and difference in strong-interaction phase between the suppressed and favoured mode, respectively, and $X$ is the $B$ final state. For $B^- \to Dh^-$ decays, the sign of $\phi_3$ is flipped. Equation~\ref{eq:amplitude} implies that the sensitivity to $\phi_3$ is approximately inversely proportional to the value of $r^X_B$, which is around 0.1 for $B^+ \to DK^+$ decays and 0.005 for $B^+ \to D\pi^+$ decays~\cite{hflav}. Hence, the sensitivity to $\phi_3$ in $B^+ \to D\pi^+$ decays is considerably worse than that in $B^+ \to DK^+$ decays, despite larger signal yields.

This paper presents a determination of $\phi_3$ from a combination of measurements using samples of up to 711~fb$^{-1}$ from the Belle experiment and up to 362~fb$^{-1}$ from the Belle~II experiment using inputs from $B^+\to DK^+, B^+\to D\pi^+$, and $B^+ \to D^{*}(\to D\pi^0, D\gamma)K^+$ decays. We use the values and correlations, when available, of the observables reported for these decays as inputs, incorporating them into a likelihood based on their relationships with $\phi_3$ and the hadronic parameters. The angle $\phi_3$ and the hadronic parameters are obtained by maximizing the likelihood and using a frequentist technique based on the Feldman-Cousins method~\cite{feldman} for the construction of confidence regions. The impact of assumptions on the unknown correlations and the relative contributions of each input to the final results are discussed. The rest of the paper is organized as follows. Section~\ref{sec:AllMethods} describes the $\phi_3$ extraction methods for the final states used in this combination.
In Section~\ref{sec:BInputs}, we present the Belle and Belle~II results entering the combination, and we list the necessary additional inputs in Section~\ref{sec:auxilaryinputs}.
Section~\ref{sec:stat} briefly outlines the statistical treatment. In Section~\ref{sec:results}, we show the results and discuss their interpretation. A summary is presented in Section~\ref{sec:summary}.

\section{Methods to obtain $\phi_{3}$}
\label{sec:AllMethods}
 
We use four methods that directly determine $\phi_3$.
We provide a brief overview of each, illustrating the relevant observables in Section~\ref{sec:BInputs}, and details of their dependences on $\phi_3$, $r_{B}^{D^{(\ast)}h}$, and $\delta_B^{D^{(\ast)}h}$ in Appendix~\ref{app:equations}.
\par The Gronau-London-Wyler (GLW) method uses the decays of $D$ mesons to $\CP$ eigenstates, such as the $\CP$-even ($\CP+$) decay $D \to K^-K^+$ and the $\CP$-odd ($\CP-$) decay $D \to \KS \piz~$\cite{glw,glw2}. 
%In the $\CP$-even decay $D \to K^-K^+$, the tiny direct $\CP$ violation of $D$ decay is ignored at the current experiment precision~\cite{LHCb:2019hro}.
The Atwood-Dunietz-Soni (ADS) method uses final states, such as $D \to K^\pm\pi^\mp$ , in which the interference between suppressed $B$ decay followed by the Cabibbo-allowed $D$ decay with the favoured $B$ decay followed by the doubly Cabibbo-suppressed $D$ decay generate large $C\!P$ asymmetries~\cite{ads1,ads2}.
%The Atwood-Dunietz-Soni (ADS) method uses \comment{final states, such as $D \to K^\pm\pi^\mp$}, in which the interference between the Cabibbo-allowed \comment{(suppressed)} and doubly Cabibbo-suppressed \comment{(favoured)} modes in both $B$ and $D$ decays generate large $\CP$ asymmetries~\cite{ads1,ads2}. 
In this method, an additional dependence on the properties of the $D$ decay is introduced through the ratio of suppressed and favoured $D$ decay amplitudes, $r_D$, and their strong-interaction phase difference, $\delta_D$. The ADS method has been extended to multibody $D$ decays, such as $D \to K^+\pi^-\piz$, where an extra coherence factor $\kappa_{D}$ is introduced to account for the dilution from the inseparable multiple interfering amplitudes integrated over the $D$-decay phase space (Dalitz plot)~\cite{nayak}.  All these parameters are measured independently and are auxiliary inputs in our combination. 

The relevant physics observables are $\CP$-violating decay-rate asymmetries,
% \begin{equation}
%   \mathcal{A}_{C\!P\pm} \equiv \frac{
%     \mathcal{B}(\PB^{-} \to \PD_{C\!P\pm} \PK^{-}) - \mathcal{B}(\PB^{+} \to \PD_{C\!P\pm} \PK^{+})
%   }{\mathcal{B}(\PB^{-} \to \PD_{C\!P\pm} \PK^{-}) + \mathcal{B}(\PB^{+} \to \PD_{C\!P\pm} \PK^{+})},
% \label{eq:acp_glw}
% \end{equation}
\begin{equation}
A_{ f} = \dfrac{\mathcal{B}(B^{-} \rightarrow \PD_{f}K^{-}) - \mathcal{B}(B^{+} \rightarrow \PD_{\bar{f}}K^{+})}{\mathcal{B}(B^{-} \rightarrow \PD_{f}K^{-}) + \mathcal{B}(B^{+} \rightarrow \PD_{\bar{f}}K^{+})},
\label{eq:acp_ads}
\end{equation}
and $\CP$-averaged decay-rate ratios, 
\begin{equation}
    R_{ f} = \dfrac{\mathcal{B}(B^{-} \rightarrow \PD_{f}K^{-}) + \mathcal{B}(B^{+} \rightarrow \PD_{\bar{f}}K^{+})}{\mathcal{B}(B^{-} \rightarrow \PD_{\bar{f}}h^{-}) + \mathcal{B}(B^{+} \rightarrow \PD_{f}h^{+})},
\label{eq:r_ads}
\end{equation}
where $f$ indicates that the $D$ meson is reconstructed in a $\CP$ eigenstate (noneigenstate) and $h$ is a pion (kaon) and $\bar{f}$ indicates the same $\CP$ (flavour conjugate) final state for the GLW (ADS) measurements.
%where $f$ represents $\CP$ (non-$\CP$) eigenstates of $D$ mesons and $h$ is a pion (kaon) for the GLW (ADS) measurements and $\bar{f}$ represents the same $\CP$ (flavour) for the GLW (ADS) measurements. 
The double ratio, for the GLW measurements, is defined as
%where $f$ represents flavour decay modes, and the double ratio for GLW measurements 
\begin{equation}
  \mathcal{R}_{C\!P\pm} \approx \frac{R_{C\!P\pm}}{R_{\rm flav}},
\label{eq:r_glw}
\end{equation}
where $R_{C\!P\pm}$ or $R_{\rm flav}$ results from specializing Equation~\ref{eq:r_ads}  for a $\CP\pm$ or flavour-specific final state, respectively. The use of double ratios helps to reduce the systematic uncertainties from branching fractions and reconstruction efficiencies of different $D$ channels appearing in the numerator and denominator of Equation~\ref{eq:r_ads}.
% where 
% \begin{equation}
%     R^{\pm}_{K/\pi} \equiv \frac{\mathcal{B}(B^{-} \rightarrow D_{\CP} K^-)+\mathcal{B}(B^{+} \rightarrow D_{\CP} K^+)}{\mathcal{B}(B^{-} \rightarrow D_{\CP} \pi^-)+\mathcal{B}(B^{+} \rightarrow D_{\CP} \pi^+)},
% \end{equation}
% and 
% \begin{equation}
%     R_{K/\pi} \equiv \frac{\mathcal{B}(B^{-} \rightarrow D_{f} K^-)+\mathcal{B}(B^{+} \rightarrow D_{\bar{f}} K^+)}{\mathcal{B}(B^{-} \rightarrow D_{f} \pi^-)+\mathcal{B}(B^{+} \rightarrow D_{\bar{f}} \pi^+)}.
% \end{equation}
Equation~\ref{eq:r_glw} is exact in the limit at which the Cabibbo-suppressed contributions to the $B^{-} \rightarrow D_{f} \pi^-$ decay amplitudes completely vanish, as detailed in Ref.~\cite{BaBar:2010hvw}.
These asymmetries and ratios are directly related to $\phi_3$ and other parameters through terms proportional to $\sin{\delta}\sin{\phi_3}$ and $\cos{\delta}\cos{\phi_3}$, respectively, as described in Appendix~\ref{app:equations}.

% \begin{equation}
% A_{\rm ADS} = \dfrac{\mathcal{B}(B^{-} \rightarrow \PD_{f}K^{-}) - \mathcal{B}(B^{+} \rightarrow \PD_{\bar{f}}K^{+})}{\mathcal{B}(B^{-} \rightarrow \PD_{f}K^{-}) + \mathcal{B}(B^{+} \rightarrow \PD_{\bar{f}}K^{+})}
% \label{eq:acp_ads}
% \end{equation}
% and the $\CP$-averaged ratio is
% \begin{equation}
%     R_{\rm ADS} = \dfrac{\mathcal{B}(B^{-} \rightarrow \PD_{f}K^{-}) + \mathcal{B}(B^{+} \rightarrow \PD_{\bar{f}}K^{+})}{\mathcal{B}(B^{-} \rightarrow \PD_{\bar{f}}K^{-}) + \mathcal{B}(B^{+} \rightarrow \PD_{f}K^{+})}.
% \label{eq:r_ads}
% \end{equation}

The Grossman-Ligeti-Soffer (GLS) method exploits singly Cabibbo-suppressed decays $D\to\KS K^\pm\pi^\mp$~\cite{glsmethod}. These two different processes are labeled as ``same-sign (SS)'' and ``opposite-sign (OS)'', according to the relationship between the charges of the parent $B$ meson and the $K$ meson from the $D$ decay. The observables include $\CP$ asymmetries of $B^+\to DK^+$ and $B^+ \to D\pi^+$ decays for each process, the $\CP$-averaged ratio of the $B^+\to DK^+$ branching-fraction relative to that of $B^+ \to D\pi^+$ for each process, and an additional ratio of the SS branching-fraction relative to that of OS for $B^+ \to D\pi^+$. This method requires information about the properties of the $D$ decay, which is encapsulated in the values of $r_D$, $\delta_D$, and $\kappa_D$, included as auxiliary external inputs in our combination.
%which are taken from external charm factories like CLEO and BESIII.
%We take this information from external charm factories like CLEO and BESIII.

The Bondar-Poluektov-Giri-Grossman-Soffer-Zupan (BPGGSZ) method relies on self-conjugate multibody-$D$-meson decays such as $\KS h^-h^+$. This method uses two approaches, which are either dependent \cite{bpggsz1} or independent~\cite{bpggsz2,bpggsz3} of the modelling of the $D \to \KS h^-h^+$ decay amplitude. The model-dependent approach relies upon a detailed description of the intermediate-resonance structure of the $D$-decay amplitude and nonresonant contributions. The model-independent method exploits {\CP}-asymmetry measurements in disjoint regions (bins) of the Dalitz plot that can be related to $\phi_3$ using model-independent measurements of $D$-decay strong-interaction-phase parameters.  The population of candidates in the Dalitz plot depends on four variables,

\begin{align}
  x_\pm^{X} = r^X_B \cos(\delta^X_B \pm \phi_3), \label{eq:xpm}\\
  y_\pm^{X} = r^X_B \sin(\delta^X_B \pm \phi_3). \label{eq:ypm}
\end{align}
The $D \to \KS h^-h^+$ decay proceeds via several intermediate resonances, which results in a variation of the $\CP$ asymmetry over the Dalitz plot, providing the best sensitivity to $\phi_3$ among all the methods. %Although the model-independent approach of this method requires $D$-decay strong-phase parameters determined by charm factories like CLEO and BESIII, they don't go as extra parameters in the combination.
%Additional two variables $x_\xi^{D\pi}, y_\xi^{D\pi}$ are used to constrain the hadronic parameters, which are defined as

%The GLW/ADS and BPGGSZ formalisms can also be extended to multibody $B$ decays by including a coherence factor, $\kappa_{B}$, that accounts for dilution from interference between competing amplitudes. % Currently we don't have multibody $B$ decays, so maybe no need for now.

Subleading effects from \(D^0 - \overline{D}^0\) mixing can impact the determination of \(\phi_3\)~\cite{mixing}. They are accounted for in this combination only for the ADS channels, where \(D^0 - \overline{D}^0\) mixing contributes at leading order in the relations between \(\phi_3\), other parameters, and the ADS observables \(R_{\rm ADS}\) and \(A_{\rm ADS}\) (see Ref.~\cite{mixing} and Equations~\ref{eq:ads1} in Appendix~\ref{app:equations}). The magnitude of the effect is inversely proportional to \(r_B^X\), making it particularly significant for \(B^+ \to D \pi^+\) decays. For consistency, \(D^0 - \overline{D}^0\) mixing effects are also included for \(B^+ \to D K^+\) modes. The contribution from \(D^0 - \overline{D}^0\) mixing cancels in the \(C\!P\) asymmetries and is negligible in the ratios of the GLW observables. Charm mixing is ignored in the BPGGSZ result, as to properly account for it a new measurement would be required taking into account its effects in the determination of the $D$-decay strong-interaction-phase parameters. However, the bias from neglecting charm mixing in BPGGSZ channels is estimated to be less than \(0.2^\circ\)~\cite{mixing}, i.e., negligible compared to the expected precision of this combination. Due to the limited precision, \(D^0 - \overline{D}^0\) mixing is also neglected in the GLS results. Finally, we ignore the small effect of direct $\CP$ violation in $D$ decays~\cite{LHCb:2022lry}.

%\comment{The sub-leading effects from \(D^0 - \bar{D}^0\) mixing can impact measurements of \(\phi_3\)~\cite{mixing}. These effects are corrected for as necessary. The magnitude of the correction is inversely proportional to \(r_B^X\), making it particularly significant for \(B^+ \to D \pi^+\) decays. To ensure consistency, the correction is also applied to the corresponding \(B^+ \to D K^+\) modes. In the GLW method, these corrections cancel out in the \(CP\) asymmetries and have negligible effects in the ratios. Conversely, in the ADS method, the corrections manifest as leading-order terms in the relationship (defined in Equations~\ref{eq:ads1}) between \(\phi_3\), other parameters, and the ADS observables \(R_{\rm ADS}\) and \(A_{\rm ADS}\)~\cite{mixing}. For the BPGGSZ method, the measurements required to be re-analysed by taking into account the $D$-decay strong-interaction-phase parameters performed with mixing effect, which is not feasible at this stage. The bias from neglecting this mixing effect is estimated to be \(\leq 0.2^\circ\)~\cite{mixing} for BPGGSZ channels. In the GLS method, due to the limited statistical precision of the GLS result, the mixing effect is neglected. Therefore, corrections are applied only to the $B^+ \to Dh^+, D \to K^+\pi^-, K^+\pi^-\piz$ channels. We also ignore the small effect of direct $\CP$ violation in $D$ decays~\cite{LHCb:2022lry}.}
%%%%%%%%%%%%%%%%%%%%%%%%%%%%%%%%%%%%%%%%%%%%%%%%%%%%%%%%%%%%%%%%%%%%%%%%%%
\section{Inputs from Belle and Belle II analyses}
\label{sec:BInputs}

We summarize the measurements used as inputs in our combination in Table~\ref{tab:listofmeasurements} and briefly describe them below. The values of the observables with their uncertainties and correlations are
provided in Appendix~\ref{app:inputs}.

\begin{table}[!h]
    \centering
    \caption{Belle and Belle II measurements used for the combination, m.i.~and m.d.~stand for model-independent and model-dependent, respectively.}
    \vspace{0.5cm}
    \resizebox{1.0\textwidth}{!}{
    \begin{tabular}{l l c c c }
    \hline
    $B$ decay & $D$ decay  & Method & Data set (Belle + Belle II)[$\invfb$] & Ref. \\
    % & & & (Belle + Belle II)[$\invfb$] & \\
     \hline
    $B^+ \to Dh^+$ & $D \to \KS \pi^0, K^-K^+$ & GLW & $711+189$ & \cite{yi-analysis} \\
    $B^+ \to Dh^+$ & $D \to K^+\pi^-, K^+\pi^-\pi^0$ & ADS & $711+0$ & \cite{horii,nayak} \\
    $B^+ \to Dh^+$ & $D \to \KS K^-\pi^+$ & GLS & $711+362$& \cite{gls} \\
    $B^+ \to Dh^+$ & $D \to \KS h^-h^+$ & BPGGSZ (m.i.) & $711+128$ & \cite{niharika} \\
    $B^+ \to Dh^+$ & $D \to \KS \pi^-\pi^+\pi^0$ & BPGGSZ (m.i.) & $711+0$ & \cite{resmi1} \\
    \multirow{2}*{$B^+ \to D^* K^+$} & $\Dstar\to D\piz, D \to \KS\piz, \KS\phi, \KS\omega,$ &\multirow{2}*{GLW} & \multirow{2}*{210+0}& \multirow{2}*{\cite{glw2}}\\
         & $K^-K^+, \pi^-\pi^+$ &  & &  \\
    $B^+ \to D^* K^+$ & $\Dstar\to D\piz, D\gamma, D \to \KS \pi^-\pi^+$ & BPGGSZ (m.d.) & $605+0$& \cite{anton} \\
    \hline
    \end{tabular}}
    \label{tab:listofmeasurements}
\end{table}

\begin{itemize}

\item \textbf{\boldmath $B^+ \to Dh^+, D \to \KS \piz, K^-K^+$.} 
This GLW measurement is based on the combined $(711 + 189)~\invfb$ Belle and Belle~II data sets, and provides two $\CP$ asymmetries and two $\CP$-averaged ratios, defined in Equations~\ref{eq:acp_ads} and \ref{eq:r_glw}, obtained from a simultaneous fit to $B^+ \to Dh^+$ decays~\cite{yi-analysis}.
    
\item \textbf{\boldmath $B^+ \to Dh^+, D \to K^+\pi^-, K^+\pi^-\piz$.} 
These ADS measurements are based on the full $711~\invfb$ Belle data set. They provide observables, such as $\CP$ asymmetry and $\CP$-averaged ratio from each channel, defined in Equations~\ref{eq:acp_ads} and \ref{eq:r_ads}~\cite{horii,nayak}.
    
\item \textbf{\boldmath $B^+ \to Dh^+, D \to \KS K^-\pi^+$.} 
This GLS measurement is based on the combined Belle and Belle~II data samples, $(711 + 362)~\invfb$ and provides four $\CP$ asymmetries and three branching-fraction ratios as inputs to the combination, defined in Equations~\ref{eq:gls}. We use only the results restricted to the quasi-two-body $D\to\Kpm\Kstarmp$ region as the resulting enhanced interference improves the expected precision~\cite{gls}.

\item \textbf{\boldmath $B^+ \to Dh^+, D \to \KS h^-h^+$.} This is a model-independent BPGGSZ measurement based on a combination of Belle and Belle II data sets corresponding to $(711 + 128)~\invfb$ of integrated luminosity~\cite{niharika}. The variables, defined in Equations~\ref{eq:xpm} and \ref{eq:ypm}, are obtained from a simultaneous fit to the Dalitz plots of $D \to \KS h^-h^+$ decays.  
In this measurement, a parametrization was adopted, which exploits  the common dependence on $\phi_3$ in $B^+\to DK^+$ and $B^+\to D\pi^+$ decays by introducing a single complex variable~\cite{GarraTico:2018nng, GarraTico:2019gdj} 
\begin{equation}
    \xi^{D\pi} = \left(\frac{r_B^{D\pi}}{r_B^{DK}}\right)e^{i\left(\delta_B^{D\pi}-\delta_{B}^{DK}\right)} \;.
    \label{eq:xi_param}
\end{equation}
The resulting input observables for $B^+ \to D\pip$ decays are defined as $x_{\xi}^{D\pi}\equiv\mathrm{Re}\left(\xi^{D\pi}\right)$ and $y_{\xi}^{D\pi}\equiv\mathrm{Im}\left(\xi^{D\pi}\right)$. The analogous observables, for $B^+ \to D\pi^+$ decays defined in Equations~\ref{eq:xpm} and \ref{eq:ypm}, are written in terms of $B^+ \to DK^+$ observables as
\begin{equation}
    x_{\pm}^{D\pi} = x_{\xi}^{D\pi}x_{\pm}^{DK}-y_{\xi}^{D\pi}y_{\pm}^{DK}\;,\;\;\; y_{\pm}^{D\pi} = x_{\xi}^{D\pi}y_{\pm}^{DK}+y_{\xi}^{D\pi}x_{\pm}^{DK} \;.
\end{equation}
Hence, this measurement provides the following six input observables for the combination: $x_\pm^{DK}, y_\pm^{DK}, x_\xi^{D\pi}$, and $y_\xi^{D\pi}$.

\item \textbf{\boldmath $B^+ \to Dh^+, D \to \KS \pi^-\pi^+\pi^0$.}
%We take the results from 
This is a model-independent BPGGSZ measurement performed on the full Belle data set, corresponding to an integrated luminosity of $711~\invfb$~\cite{resmi1}.
The variables, defined in Equations~\ref{eq:xpm} and \ref{eq:ypm}, are obtained using a fit to the Dalitz plot of $D \to \KS \pi^-\pi^+\piz$ decays.

\item \textbf{\boldmath $B^+ \to \Dstar K^+, \Dstar\to D\piz, D\to \KS\piz, \KS\phi, \KS\omega, K^-K^+, \pi^-\pi^+$.} 
This GLW measurement is based on a $210~\invfb$ subset of Belle data~\cite{glw2}. The input observables are defined in Equations~\ref{eq:acp_ads} and \ref{eq:r_glw}. 

\item \textbf{\boldmath $B^+ \to \Dstar K^+, \Dstar\to D\piz, D\gamma, D\to \KS\pi^-\pi^+$.} 
This is a model-dependent BPGGSZ measurement based on  a $605~\invfb$ subset of Belle data~\cite{anton}. The input observables are defined in Equations~\ref{eq:xpm} and \ref{eq:ypm}. 
\end{itemize}

We do not use inputs from $B^0 \to D^{(*)}h^{(*)}$ decays~\cite{negishi1,seema,negishi2} in the combination because, due to their limited precision and their dependence on additional external parameters, they would have negligible impact on the determination of $\phi_3$. 
%%%%%%%%%%%%%%%%%%%%%%%%%%%%%%%%%%%%%%%%%%%%%%%%%%%%%%%%%%%%%%%%%%%%%%%%%%
\section{Auxiliary inputs}
\label{sec:auxilaryinputs}
Several auxiliary inputs are needed to constrain the $D$-decay parameters to extract $\phi_3$.  These are summarized in Table~\ref{tab:ExternalInputsvalues} and briefly described below.  
Correlations between inputs are reported in Appendix~\ref{app:charminputs}.
\begin{table}[!h]
    \centering
    \caption{Auxiliary input observables and their values used in the $\phi_3$ combination.}
    \vspace{0.5cm}
    \resizebox{1.0\textwidth}{!}{
    \renewcommand*{\arraystretch}{1.15}
    \begin{tabular}{l c c c c}
    \hline
    Decay & Observable  & Value & Source & Reference\\
    \hline
    \multirow{4}*{$D\to K^+\pi^-$} & $R_D^{K\pi}$ & $(3.44 \pm 0.02)\times 10^{-3}$ & \multirow{2}*{HFLAV} & \multirow{2}*{\cite{hflav}}\\
     &$\delta_D^{K\pi}$ & $(191.7 \pm 3.7)^\circ $ &  & \\
     \cdashline{2-5}
     &$r_D^{K\pi}\cos(\delta_D^{K\pi})$ & $-0.0562 \pm 0.0081$ & \multirow{2}*{BESIII} & \multirow{2}*{\cite{CharmKpiBESIII}}\\
     &$r_D^{K\pi}\sin(\delta_D^{K\pi})$ &  $-0.011 \pm 0.012$ &  & \\
     \hline
    \multirow{3}*{$D\to K^+\pi^-\piz$}& $r_D^{K\pi\piz}$ & $0.0441 \pm 0.0011$ & \multirow{3}*{CLEO + LHCb + BESIII} & \multirow{3}*{\cite{CharmKpipi0BESIII}}\\
    & $\kappa_D^{K\pi\piz}$ & $0.79 \pm 0.04$ &  & \\
    & $\delta_D^{K\pi\piz}$ & $(196 \pm 11)^\circ$ &  & \\
    \hline
    \multirow{2}*{\(D^0-\overline{D}^0\) mixing} & $x_D$ & $(0.407 \pm 0.044)\%$ & \multirow{2}*{HFLAV} & \multirow{2}*{\cite{hflav}}\\
     & $y_D$ & $(0.647 \pm 0.024)\%$ & & \\
    \hline
    \multirow{4}*{$D\to \KS K^-\pi^+$} & $(r_D^{\KS K\pi})^2$ & $0.356 \pm 0.034$ & \multirow{3}*{CLEO} & \multirow{3}*{\cite{glsCharmCLEO}}\\
     & $ \kappa_D^{\KS K\pi}$ & $0.94 \pm 0.12$ &  & \\
     & $ \delta_D^{\KS K\pi}$ & $(-16.6 \pm 18.4)^\circ$ &  & \\
     \cdashline{2-5}
     & $(r_D^{\KS K\pi})^2$ & $0.370 \pm 0.003$ & LHCb & \cite{glsCharmLHCb}\\
     \hline
    $B^+ \to Dh^+$ & $R_{\textrm{GLS}}$ & 0.0789$\pm$0.0027 & PDG & \cite{pdg}\\
    \hline
    \end{tabular}}
    \label{tab:ExternalInputsvalues}
\end{table}

\begin{itemize}
\item \textbf{\boldmath Input for $D \to K^+\pi^-$ decays.}
The ADS measurement requires the ratio $r_{D}^{K\pi}$ and strong-interaction phase difference $\delta_{D}^{K\pi}$ between favoured $D \to K^-\pi^+$ and suppressed $D \to K^+\pi^-$ decays to constrain the properties of the charm system. We take these from the HFLAV global fit to measurements of $\CP$-violation and mixing in the $D^0-\overline{D}^0$ system~\cite{hflav}. The value of $\delta_D^{K\pi}$ is shifted by $180\degrees$ compared to Ref.~\cite{hflav} to match the phase convention adopted in this work. The parameter $R_D^{K\pi}$ is the square of the amplitude ratio $r_D^{K\pi}$. We also include a measurement from BESIII~\cite{CharmKpiBESIII} performed on a 2.93~$\invfb$ $\psi(3770)$ data set, which is not included in Ref.~\cite{hflav}.

\item \textbf{\boldmath Input for $D \to K^+\pi^-\piz$ decays.}
The ADS measurement with $D \to K^+\pi^-\piz$ decays requires
knowledge of hadronic parameters describing the $D$ decay. These are the amplitude ratio $r_D^{K\pi\piz}$, strong-interaction phase difference $\delta_D^{K\pi\piz}$, and coherence factor $\kappa_D^{K\pi\piz}$.
We take the combined result of BESIII, CLEO, and LHCb from Ref.~\cite{CharmKpipi0BESIII}.

\item \textbf{\boldmath \(D^0 - \overline{D}^0\) mixing parameters.} The ADS measurements require the charm mixing parameters \(x_D\) and \(y_D\) as inputs. We obtain these inputs from the HFLAV global fit to measurements of \(\CP\)-violation and mixing in the \(D^0 - \overline{D}^0\) system~\cite{hflav}. The HFLAV average of \(x_D\) is dominated by the LHCb measurement of Ref.~\cite{LHCb:2021ykz}, which uses as input the same $D$-decay strong-interaction-phase parameters as used in our BPGGSZ result. The correlation introduced by these common inputs is neglected in this combination, as the impact of the $D$-decay strong-interaction-phase parameters in the BPGGSZ result is small compared to the precision of the measurement~\cite{niharika}.
%which are currently dominated by the LHCb measurement~\cite{LHCb:2021ykz}. This LHCb measurement uses the same values of \(D\)-decay strong-interaction-phase parameters~\cite{BESIII:2020khq} that are employed in our BPGGSZ measurement~\cite{niharika}. Therefore, some correlation is expected between these and the mixing parameters. However, as the impact of the \(D\)-decay strong-interaction-phase parameters is small for our BPGGSZ measurement, as described in Ref.~\cite{niharika}, we neglect these correlations in our combination.

\item \textbf{\boldmath Input for $D \to \KS K^-\pi^+$ decays.}
The GLS measurement with $D \to \KS K^-\pi^+$ decays requires the hadronic parameters $r_D^{\KS K \pi}, \delta_D^{\KS K \pi}$, and $\kappa_D^{\KS K \pi}$. We use the measurement from CLEO~\cite{glsCharmCLEO} and the $r_D^{\KS K \pi}$ result from LHCb~\cite{glsCharmLHCb}.  In addition,  we take the ratio of branching fractions $R_{\textrm{GLS}} = \BR(B^- \to D^0K^-)/\BR(B^- \to D^0\pi^-)$ from Ref.~\cite{pdg}.

\end{itemize}

% In our study, we do not include $D^0 - \overline{D}^{0}$ mixing due to its negligible impact on $\phi_3$, with a magnitude of only $(0.7 \pm 0.7)^\circ$, when compared to the precision of current experimental measurements~\cite{mixing}.
% We also ignore the small effect of direct $\CP$ violation in $D$ decays~\cite{LHCb:2022lry}.
%\footnote{The only observed direct $\CP$ violation are for $D\to K^+K^-$ and $D\to \pi^+\pi^-$ decays~\cite{LHCb:2019hro}.}
%%%%%%%%%%%%%%%%%%%%%%%%%%%%%%%%%%%%%%%%%%%%%%%%%%%%%%%%%%%%%%%%%%%%%%%%%%
\section{Statistical treatment}
\label{sec:stat}
We determine $\phi_3$ and six hadronic parameters $r_{B}^{DK}, \delta_B^{DK}, r_{B}^{D\pi}, \delta_B^{DK}, r_{B}^{D^*K}$, and $ \delta_B^{D^*K}$ using the relations defined in Equations~\ref{eq:bpggsz},~\ref{eq:ads1},~\ref{eq:glw1}, and~\ref{eq:gls}. These relations require values of eight $D$-decay parameters and one $B$-decay ratio,
%$ R(\BR(D^0K^-)/\BR(D^0\pi^-))$. The plugged-in values of these external parameters are
summarized in Section~\ref{sec:auxilaryinputs}. 
We combine all auxiliary inputs and results from Belle and Belle~II measurements in a maximum-likelihood fit.
%The statistical methods used to calculate the best value of the physics parameters and confidence regions are discussed below.
\par We denote the set of all experimental observables as $\Vec{X}$ and underlying physics parameters including $\phi_3$ as $\Vec{\theta}$. For a particular set of observables, $\Vec{X}^{\rm obs}$, the likelihood function is defined as the product of the probability density functions (PDFs),
\begin{equation}
\mathcal{L}(\Vec{\theta} \vert \Vec{X}^{\rm obs}) = \prod_i f_i(\Vec{X}^{\rm obs}_i\vert\Vec{\theta}),
\label{eq_chap5:l}
\end{equation} 
where $ f_i(\Vec{X}^{\rm obs}_i\vert\Vec{\theta}) $ is the PDF of 
observables  $\Vec{X}^{\rm obs}_i$ for each measurement $i$. For each of the inputs, we assume the observables follow Gaussian distributions
\begin{equation}
f_i(\Vec{X}^{\rm obs}_i\vert\Vec{\theta}) \varpropto \textrm{exp} \left\{-\frac{1}{2}[ \Vec{X}_i(\Vec{\theta}) - \Vec{X}^{\rm{obs}}_i]^T V^{-1}_i[\Vec{X}_i(\Vec{\theta}) - \Vec{X}^{\rm{obs}}_i]\right\},
\label{eq_chap5:m}
\end{equation} 
where $V^{-1}_i$ is the experimental covariance matrix, which accounts for statistical and systematic uncertainties and their correlations. The correlation of systematic uncertainties within an experiment is ignored. For the case where the uncertainties are asymmetric, we
symmetrize them, without changing the central value, by substituting the standard deviation of the distribution observed in simulated experiments generated using the asymmetric Gaussian likelihood function.
We estimate $\Vec{\theta}$ by minimizing a $\chi^{2}$-like quantity defined as $\chi^{2}(\Vec{\theta}\vert \Vec{X}^{\rm obs}) = -2\,{\rm ln}\mathcal{L}(\Vec{\theta}\vert \Vec{X}^{\rm obs})$. The best-fit value is given by the global minimum of the $\chi^2$ function, $\chi^2(\Vec{\theta}_{\rm min})$.

To estimate the confidence level (CL) for each parameter, we use the test statistic defined as $\Delta \chi^2 = \chi^2(\Vec{\theta}'_{\rm min}) - \chi^2(\Vec{\theta}_{\rm min})$, where $\chi^2(\Vec{\theta}'_{\rm min})$ is the $\chi^2$ function at the $\Vec{\theta}'$ value of the parameter. 
We generate simulated experiments with parameters $\Vec{\theta}$ set to $\Vec{\theta}'_{\rm min}$ and calculate $\Delta \chi^{2'}$ by replacing $\Vec{X}_{\rm obs}$ with the simulated experiments and minimising with respect to $\Vec{\theta}$. The value of 1 $-$ CL is calculated as the fraction of the simulated experiments
that have larger $\Delta \chi^{2'}$ $(\Delta \chi^2 < \Delta \chi^{2'})$ than the measured data.
%which perform worse ($\Delta \chi^2 < \Delta \chi^{2'}$) than the measured data~\cite{gammacombo,Aaij_2016}.}
%The value of 1 $-$ CL is calculated using an approximate likelihood-ratio ordering~\cite{feldman} based on simulated experiments sampled from the PDF, taking into account the presence of nuisance parameters, whose true values are assumed to be the best-fit values from data~\cite{gammacombo}. 
This approach is known as the \textsc{Plugin} method.

%%%%%%%%%%%%%%%%%%%%%%%%%%%%%%%%%%%%%%%%%%%%%%%%%%%%%%%%%%%%%%%%%%%%%%%%%%
\section{Results}
\label{sec:results}

\begin{figure}[htp]
  \centering
  \begin{overpic}[width=0.49\textwidth]{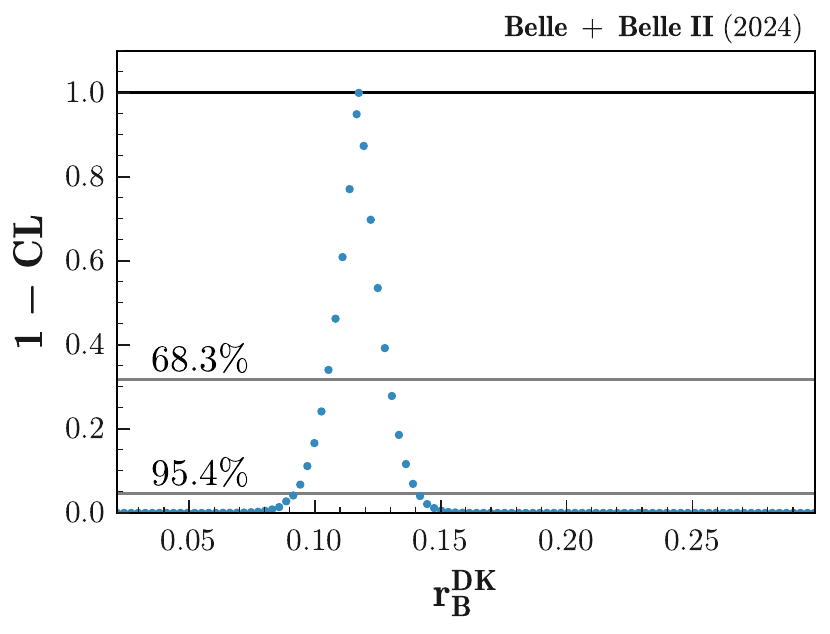}\put(75,45){(a)}\end{overpic}
  \begin{overpic}[width=0.49\textwidth]{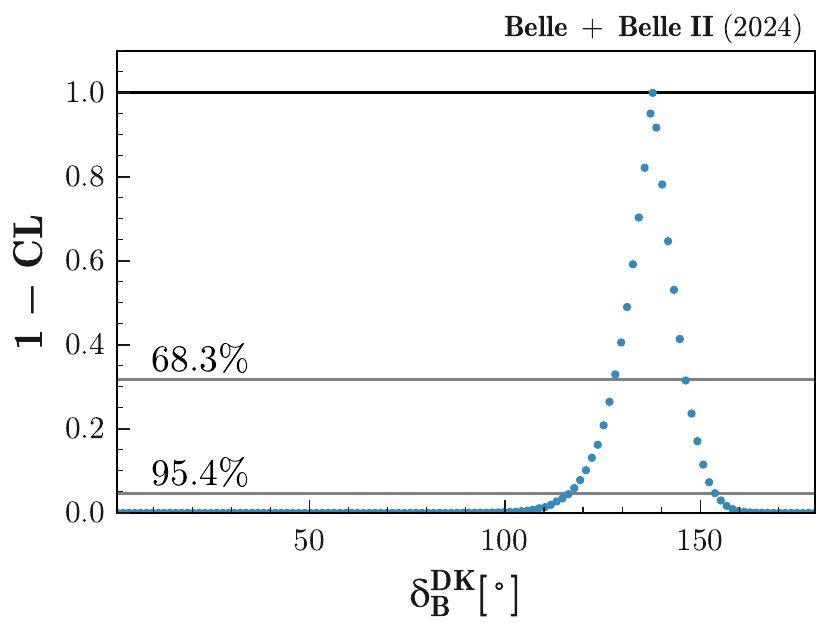}\put(85,45){(b)}\end{overpic}  
  \begin{overpic}[width=0.49\textwidth]{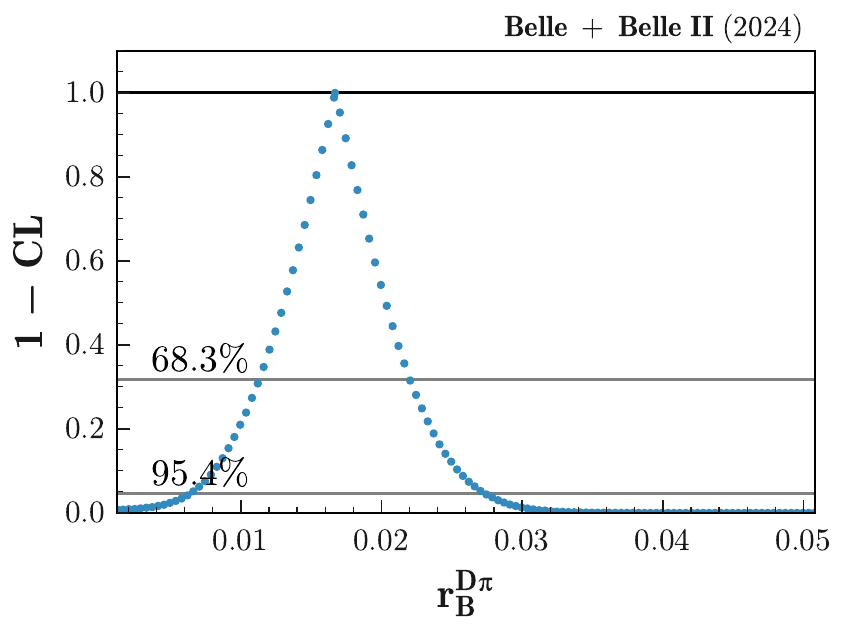}\put(75,45){(c)}\end{overpic}
  \begin{overpic}[width=0.49\textwidth]{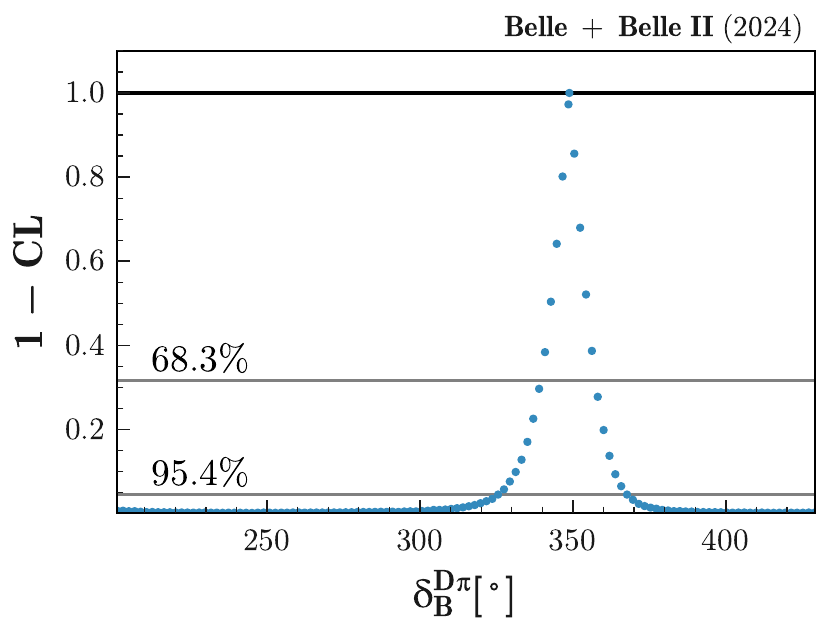}\put(75,45){(d)}\end{overpic}
  \begin{overpic}[width=0.49\textwidth]{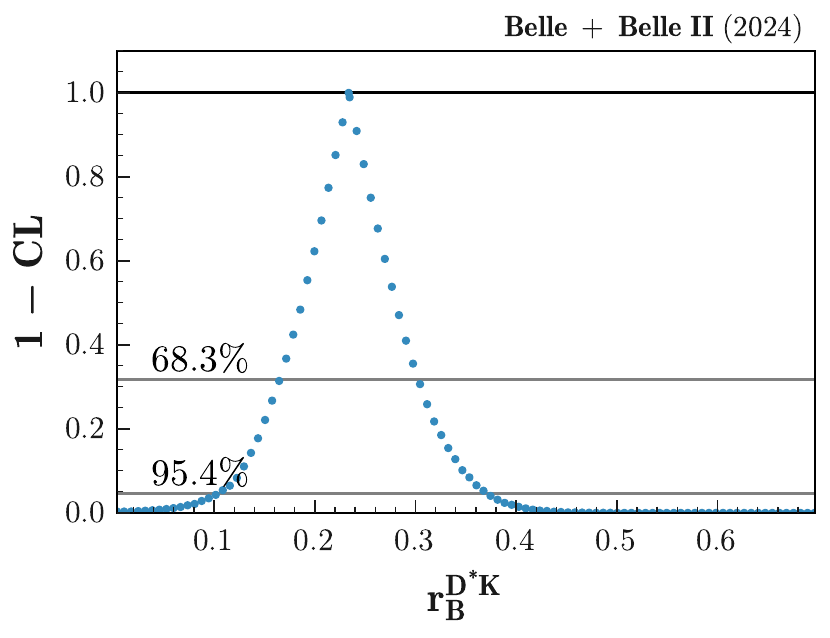}\put(75,45){(e)}\end{overpic}
  \begin{overpic}[width=0.49\textwidth]{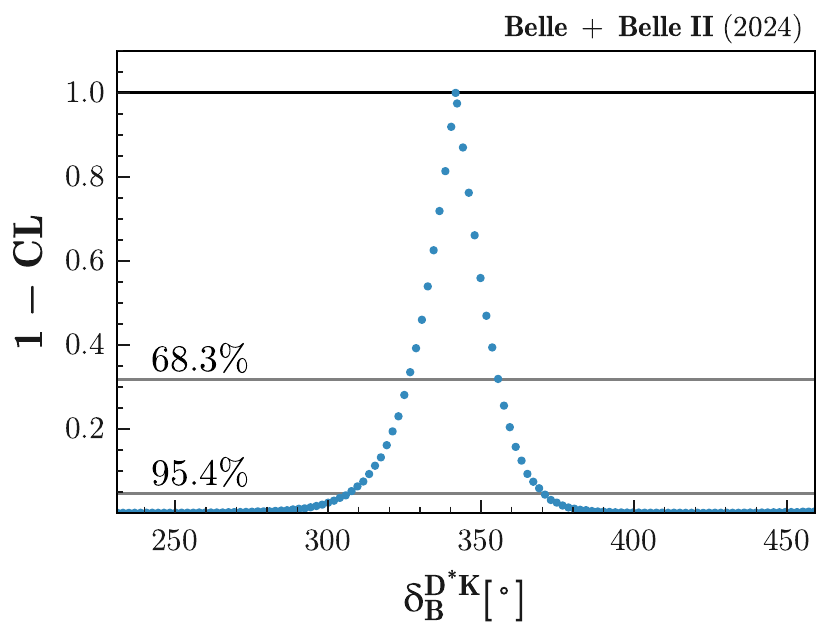}\put(75,45){(f)}\end{overpic}  
  \begin{overpic}[width=0.49\textwidth]{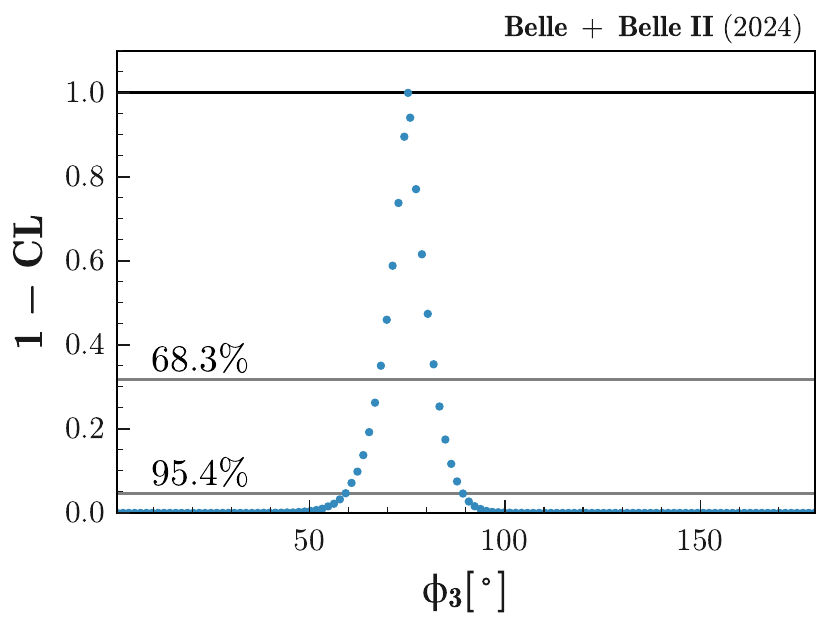}\put(75,45){(g)}\end{overpic}  
  \caption{1 $-$ CL distributions as functions of (a) $r_B^{DK}$, (b) $\delta_{B}^{DK}$, (c) $r_B^{D\pi}$, (d) $\delta_{B}^{D\pi}$, (e) $r_B^{D^\ast K}$, (f) $\delta_{B}^{D^\ast K}$,  and (g) $\phi_{3}$.}
  \label{fig:DhDstK1DPlot}
\end{figure}

\begin{figure}[htp]
  \centering
  \begin{overpic}[width=0.49\textwidth]{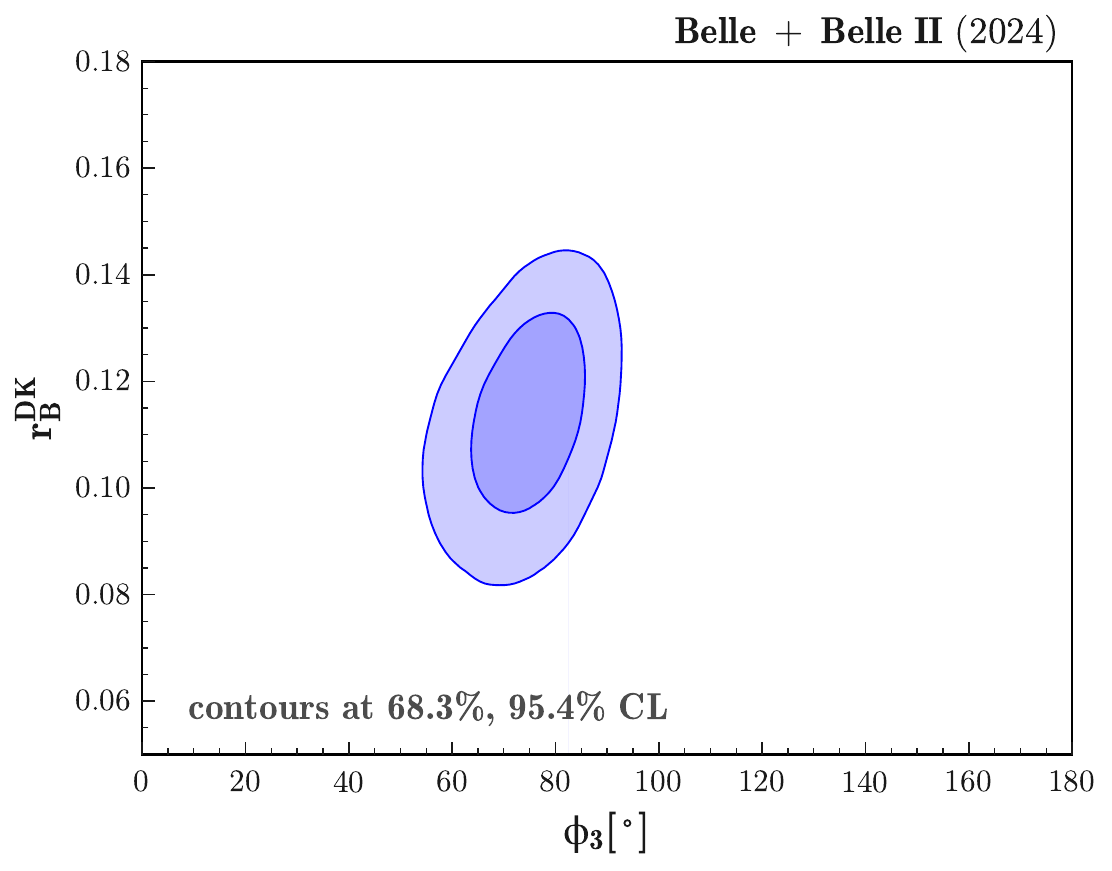}\put(30,60){(a)}\end{overpic}
  \begin{overpic}[width=0.49\textwidth]{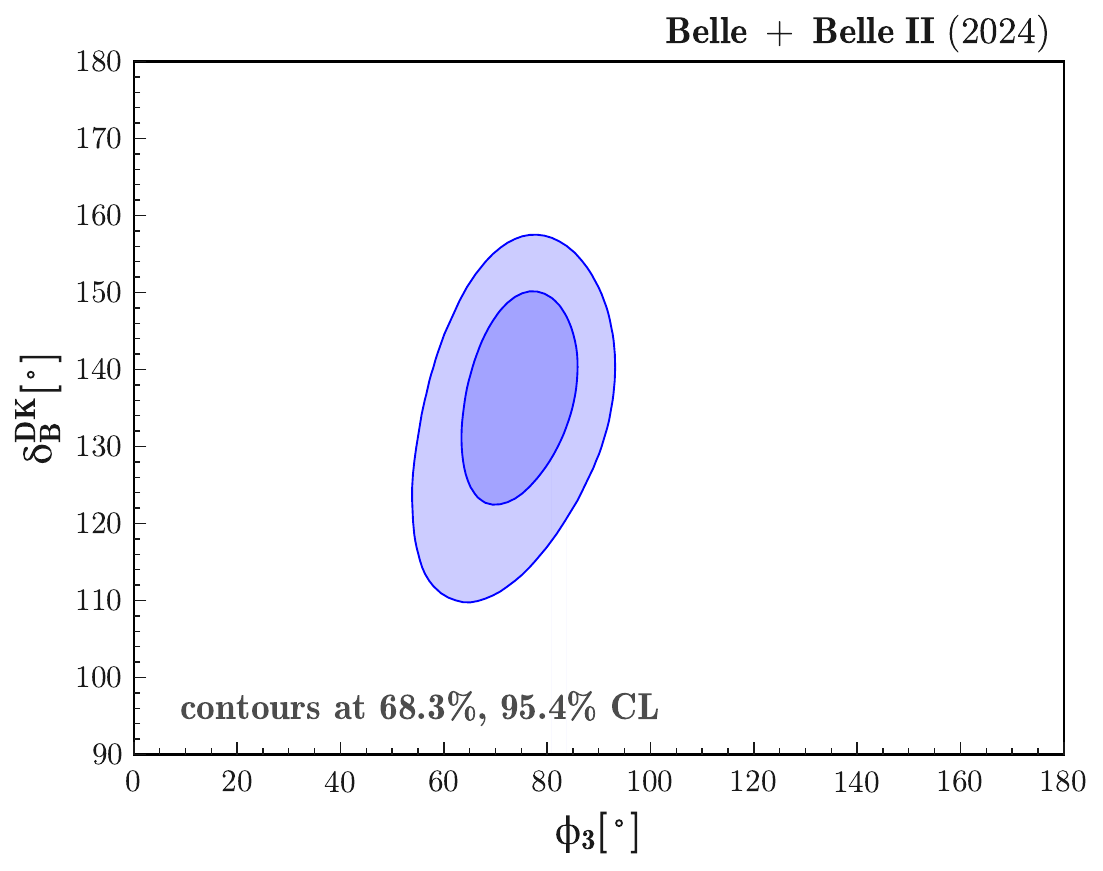}\put(30,60){(b)}\end{overpic}  
  \begin{overpic}[width=0.49\textwidth]{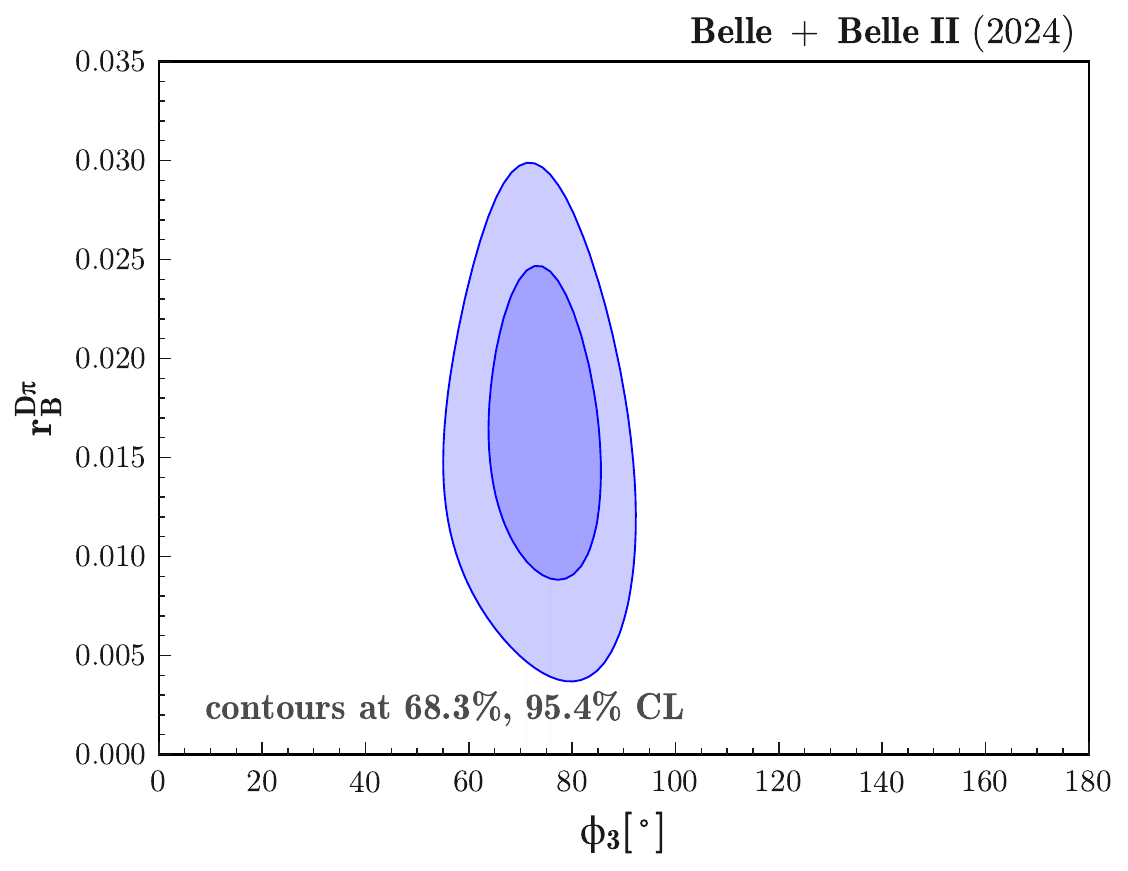}\put(30,60){(c)}\end{overpic}
  \begin{overpic}[width=0.49\textwidth]{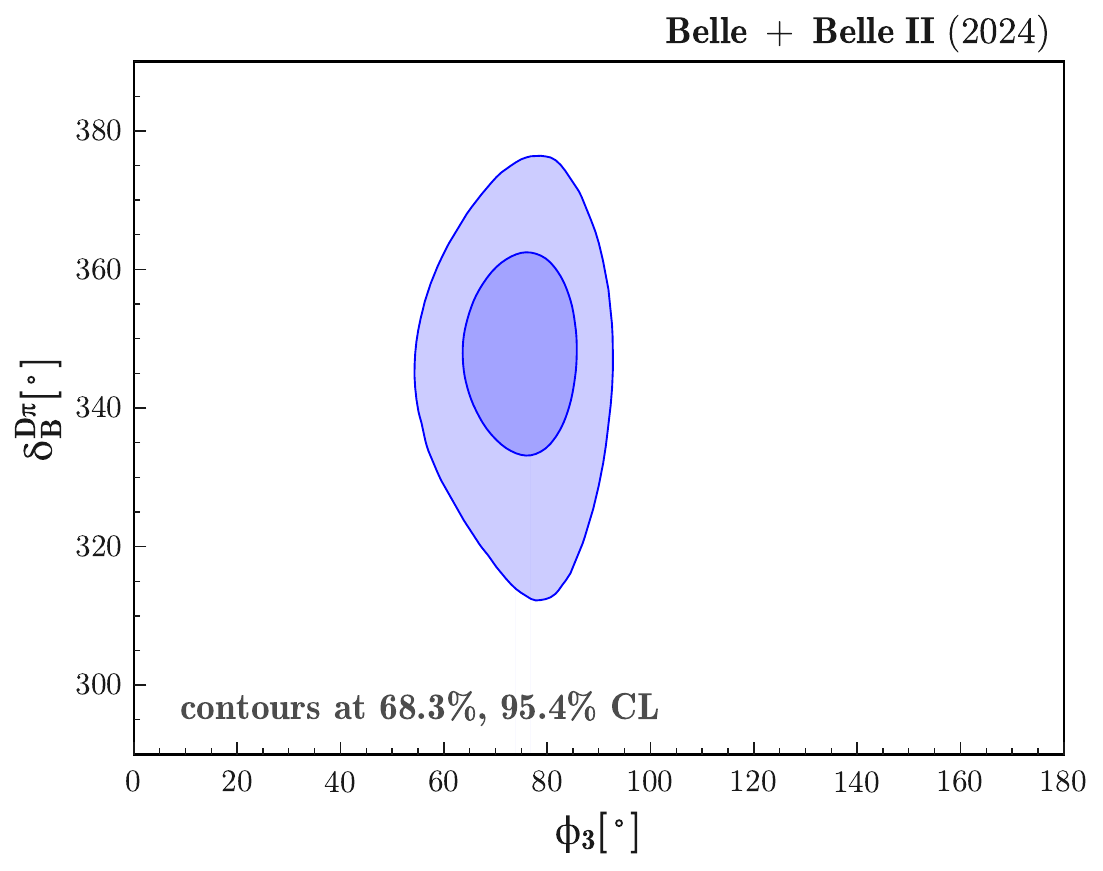}\put(30,60){(d)}\end{overpic}
  \begin{overpic}[width=0.49\textwidth]{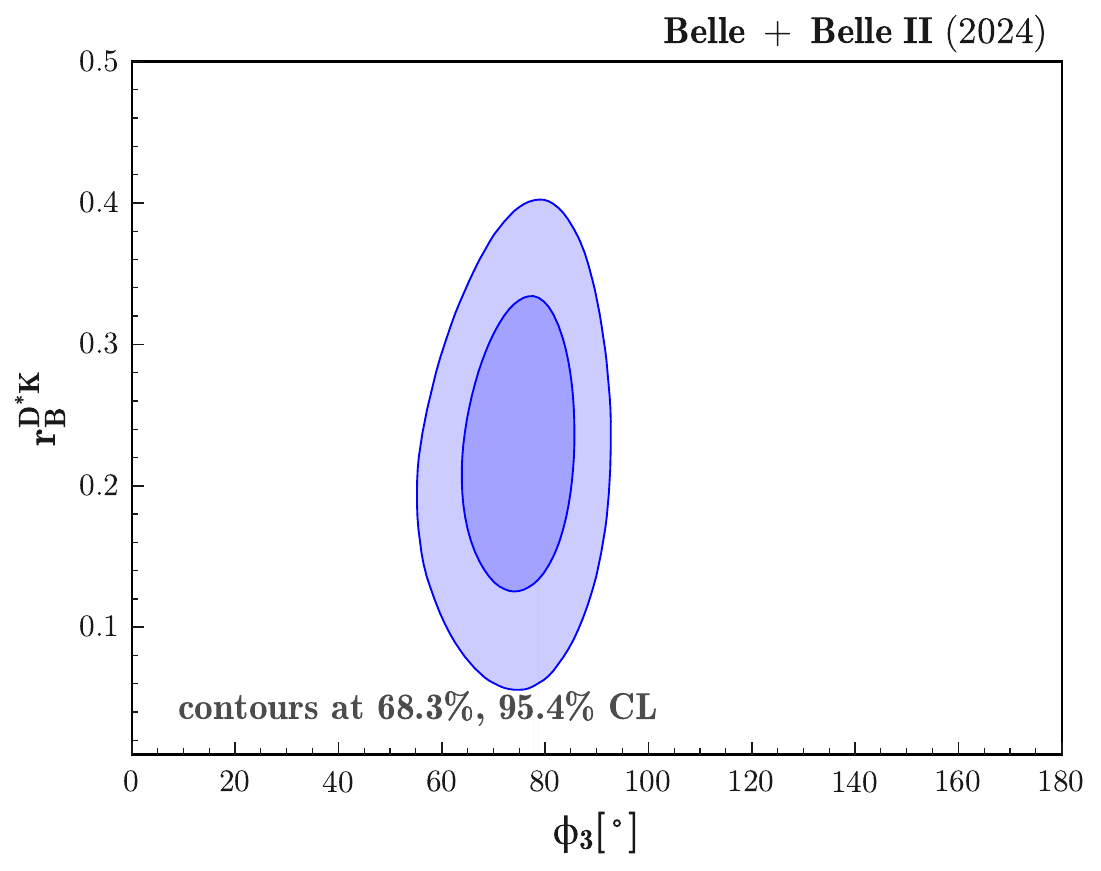}\put(30,60){(e)}\end{overpic}
  \begin{overpic}[width=0.49\textwidth]{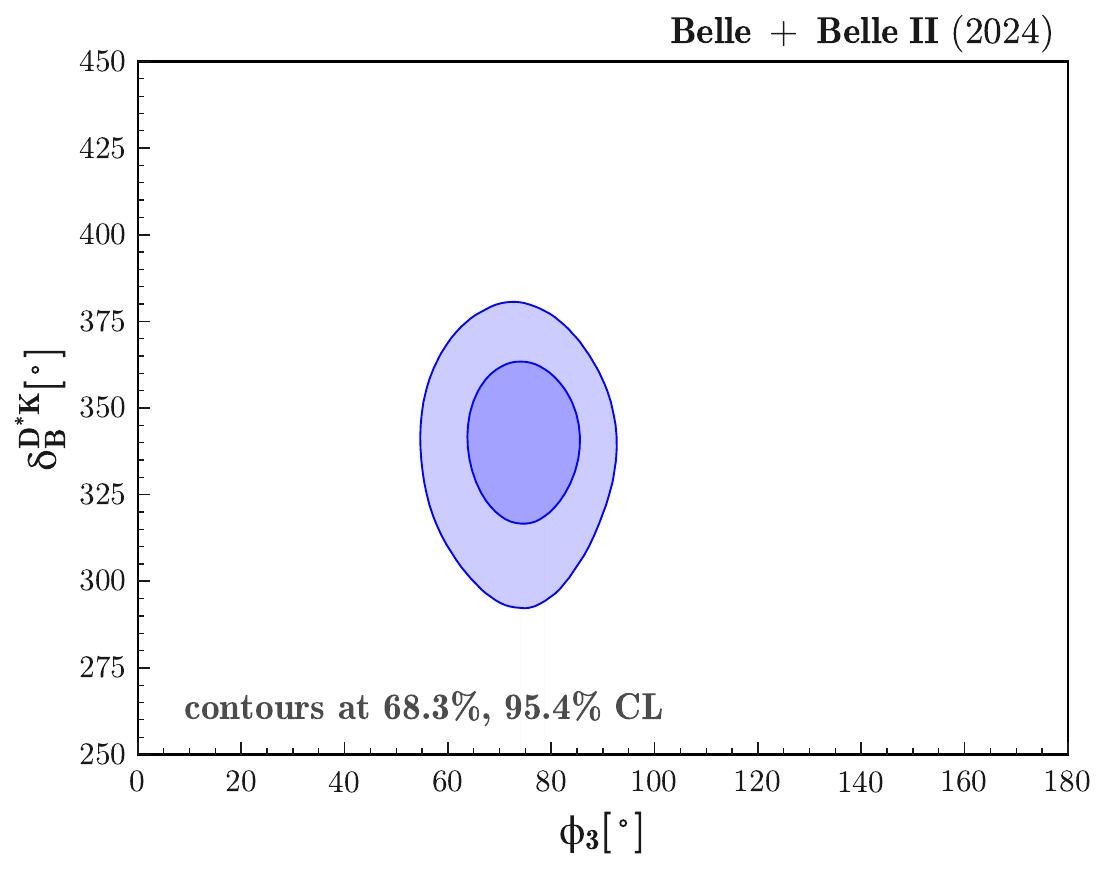}\put(30,60){(f)}\end{overpic}
  \caption{Two-dimensional confidence regions at the (inner curve) 68.3\% and (outer curve) 95.4\% confidence levels, obtained for (a) $ r_{B}^{DK}-\phi_{3} $, (b) $ \delta_{B}^{DK}-\phi_{3}$, (c) $ r_{B}^{D\pi}-\phi_{3} $, (d) $ \delta_{B}^{D\pi}-\phi_{3} $, (e) $ r_{B}^{D^\ast K}-\phi_{3} $, and (f) $ \delta_{B}^{D^\ast K}-\phi_{3} $.}
  \label{fig:DhDstK2DPlot}
\end{figure}

We combine 59 input observables from the measurements listed in Tables~\ref{tab:listofmeasurements} and \ref{tab:ExternalInputsvalues} to determine $\phi_3$ and the six $B$-decay hadronic parameters $r_B^{D^{(\ast)}h}$ and $\delta_B^{D^{(\ast)}h}$.
The fit has a total of 18 free parameters, including eight $D$-decay hadronic parameters.
We obtain $\phi_{3} = (75.2^{+7.1}_{-7.5})^{\circ}$, where the uncertainties include the statistical and systematic contributions from all inputs. The results for other parameters are summarized in Table~\ref{tab:DhDstK1DResult1}, where we report central values and confidence intervals at 68.3\% and 95.4\% probability.
We show the $\chi^2$ values for each measurement in Appendix~\ref{app:chisq}. We also show the pull distribution in Appendix~\ref{app:pull}, which is defined as $(A_{\textrm{obs}} - A_{\textrm{fit}})/\sigma({\textrm{obs}})$, where $A_{\textrm{obs}}$ and $A_{\textrm{fit}}$ are the input value and the best-fit value, respectively, and $\sigma({\textrm{obs}})$ is the measurement uncertainty.
Table \ref{tab:DhDstK1DCorr} gives the combined statistical and systematic correlation matrix for these parameters.
The goodness of the fit calculated from the fraction of simulated experiments,
generated from the best-fit point, which have a $\chi^2$ larger than that found in the data is $p=(55.4\pm 0.2)\%$.
We perform a one-dimensional profile-likelihood scan for $\phi_{3}$, the strong-interaction phase $\delta_{B}^{D^{(\ast)}h}$, and the amplitude ratio $r_{B}^{D^{(\ast)}h}$. 
Figure~\ref{fig:DhDstK1DPlot} shows the 1 $-$ CL distributions as function of the scanned parameters. 
We also perform the two-dimensional profile-likelihood scan for ($\phi_{3}$, $r_{B}^{D^{(\ast)}h}$) and ($\phi_{3}$, $\delta_{B}^{D^{(\ast)}h}$). The corresponding confidence regions are shown in Figure~\ref{fig:DhDstK2DPlot}.

\begin{table}[!h]
\centering
    \caption{Combination results: best-fit values and 68.3\% and 95.4\% confidence intervals.}
    \vspace{0.5cm}
    \setlength{\tabcolsep}{1mm}{

    \resizebox{1.0\textwidth}{!}{
    \begin{tabular}{c c c c c c c c }
    \hline
    Parameters & $\phi_3 (^\circ)$ & $r_B^{DK}$  & $\delta_B^{DK} (^\circ)$  & $r_B^{D\pi}$  & $\delta_B^{D\pi} (^\circ)$ & $r_B^{D^*K}$ & $\delta_B^{D^*K} (^\circ)$ \\
    \hline
    Best-fit value & 75.2 & 0.115 & 137.8 & 0.0165 & 347.0 & 0.229 & 342\\
    68.3\% interval & [67.7, 82.3] & [0.102, 0.127]  & [128.0, 146.3]& [0.0113, 0.0220]& [337.4, 355.7] & [0.162, 0.297] & [326,356]\\
    95.4\% interval & [59, 89] &[0.089, 0.138] &[116, 154] &[0.006, 0.027] &[322, 366] & [0.10, 0.37] & [306, 371]\\
    \hline
    \end{tabular}}}
    \label{tab:DhDstK1DResult1}
\end{table}

\begin{table}[!h]
    \centering
    \caption{Combined statistical and systematic correlations between $\phi_3$ and hadronic parameters.}
    \vspace{0.5cm}
    \begin{tabular}{c | c c c c c c c }
    \hline
     & $\phi_3$ & $r_B^{DK}$  & $\delta_B^{DK}$  & $r_B^{D\pi}$  & $\delta_B^{D\pi}$ & $r_B^{D^*K}$ & $\delta_B^{D^*K}$ \\
    \hline                  %update at 6.13
    $\phi_3$ & 1.000 & 0.364 & 0.325 & $-$0.158 &  0.005 & $\phantom{-}$0.155 & $-$0.016\\
    $r_B^{DK}$ & & 1.000 & 0.256 & $\phantom{-}$0.054 & 0.012 & $\phantom{-}$0.056 &  $-$0.006\\
    $\delta_B^{DK}$ & & & 1.000 & $\phantom{-}$0.111 & 0.105 & $\phantom{-}$0.050 & $-$0.005\\
    $r_B^{D\pi}$ & & & & $\phantom{-}$1.000 & 0.146 & $-$0.025 & $\phantom{-}$0.003\\
    $\delta_B^{D\pi}$ & & & & & 1.000 & $\phantom{-}$0.000 & $\phantom{-}$0.000\\
    $\delta_B^{D\pi}$ & & & & & 1.000 & $\phantom{-}$0.000 & $\phantom{-}$0.000\\
    $r_B^{D^*K}$ & & & & & & $\phantom{-}$1.000 & $\phantom{-}$0.168 \\
    $\delta_B^{D^*K}$ & & & & & & & $\phantom{-}$1.000 \\
    \hline
    \end{tabular}
    \label{tab:DhDstK1DCorr}
\end{table}

% \begin{table}[!h]
% \centering
%     \caption{Result of the $Dh + D^{*}K$ combination with the 68.3\%, 95.5\% and 99.7\% confidence intervals.}
%     \vspace{0.5cm}
%     \setlength{\tabcolsep}{1mm}{
%     \begin{tabular}{c c c }
%     \hline
%     Parameters  & $r_B^{D^*K}$ & $\delta_B^{D^*K} (^\circ)$ \\
%     \hline
%     \multicolumn{3}{c}{\textsc{Prob} method} \\
%     \hline              %update at 06.13
%     central value &  0.234 & 341\\
%     68.3\% interval &  [0.164, 0.304] & [327, 355]\\
%     95.5\% interval &  [0.10, 0.37] & [307, 369]\\
% %    99.7\% interval &  [0.03, 0.44] & [274, 390]\\ 
%     \hline
%     \multicolumn{3}{c}{\textsc{Plugin} method} \\
%     \hline
%     central value &  0.234 & 341\\
%     68.3\% interval &  [0.165, 0.303] & [327,355]\\
%     95.5\% interval &  [0.10, 0.37] & [307, 369]\\
% %    99.7\% interval &  [0.00, 0.44] & [273, 390]\\ 
%     \hline
%     \end{tabular}}
%     \label{tab:DhDstK1DResult2}
% \end{table}

\begin{figure}[htp]
  \centering
  \begin{overpic}[width=0.49\textwidth]{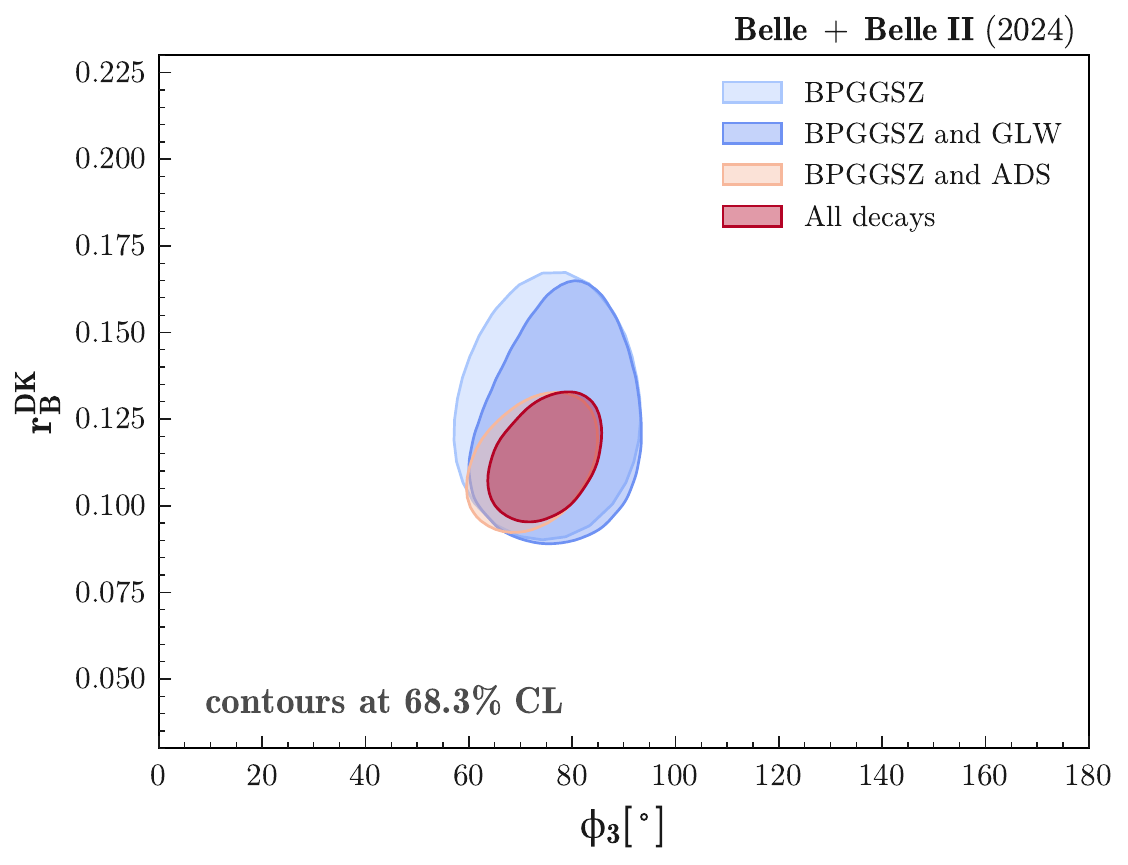}\end{overpic}
  \begin{overpic}[width=0.49\textwidth]{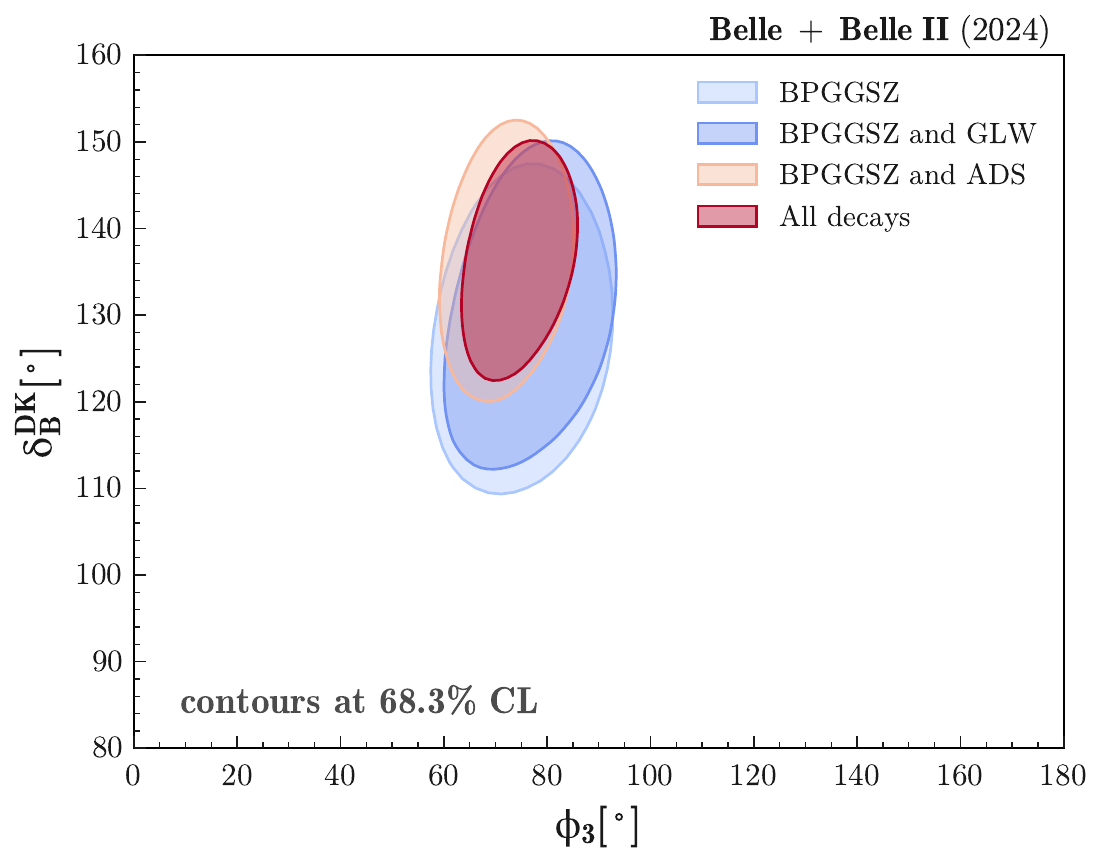}\end{overpic}  
  \caption{Two-dimensional confidence regions at the 68.3\% confidence levels, combining measurements based on various methods, for $ r_{B}^{DK}-\phi_{3} $ (left) and $ \delta_{B}^{DK}-\phi_{3}$ (right).}
  \label{fig:multicombiner}
\end{figure}

We also investigate the individual contributions of each method by presenting two-dimensional confidence regions for various configurations of combinations that include only subsets of inputs, as shown in Figure~\ref{fig:multicombiner}. The $68.3\%$ confidence intervals for these are listed in Table~\ref{tab:multicombiner}. As expected, the sensitivity is mostly dominated by the BPGGSZ measurements. Following closely are inputs from the ADS-like final states, with the GLW measurements and other results being the next most significant. We do not report the individual contributions of the ADS and GLW inputs, as they are strongly correlated with nuisance parameters obtained from the predominant BPGGSZ measurements in this combination. A comprehensive discussion of this effect can be found below.

\begin{table}[htp]
    \centering
    \caption{One-dimensional confidence intervals at the 68.3\% probability, derived from the combination of measurements from various methods. Additional uncertainties due to assumptions on the unknown correlations are not included in these intervals.}
    \vspace{0.5cm}
    %\resizebox{1.0\textwidth}{!}{
    \begin{tabular}{c c c c  }
    \hline
    Method & BPGGSZ & BPGGSZ and GLW & BPGGSZ and ADS\\
    \hline
    $\phi_3 (^\circ)$ & [65, 87] & [68, 90] & [64.2, 80.8] \\
    $r_B^{DK}$  & [0.104, 0.156] & [0.098, 0.146] & [0.100, 0.126]\\
    $\delta_B^{DK} (^\circ)$ & [118, 142] & [117, 143] & [126, 148] \\
    $r_B^{D\pi}$  & [0.0111, 0.0235] & [0.0117, 0.0237] & [0.0118, 0.0223] \\
    $\delta_B^{D\pi} (^\circ)$ & [317, 355] & [323, 357] & [337.4, 355.4]  \\    
    \hline
    \end{tabular}%}
    \label{tab:multicombiner}
\end{table}
A proper combination should include the statistical and systematic correlations between inputs.  However, correlations between input observables are not available for all modes. For example, statistical correlations are not given for Refs.~\cite{glw2} and~\cite{nayak} and systematic correlations are not provided by Refs.~\cite{glw2},~\cite{nayak},~\cite{horii},~\cite{resmi1}, and~\cite{anton}. Our choice is to set the unknown correlations to zero in the central combination and assign an additional uncertainty to the results due to this assumption. To calculate the additional uncertainty, we set the unknown statistical correlations to $\pm 0.3$ and monitor the changes in results when varying the unknown systematic correlations up to $\pm 0.9$. The choice of the value $\pm 0.3$ is based on knowledge of known correlations in other ADS and GLW-like final states, as reported in Refs.~\cite{yi-analysis} and~\cite{horii}, respectively. We examine the results when correlations are changed individually and collectively and assign the largest difference observed on the $\phi_3$ central values with respect to the nominal result as an additional uncertainty. For $\phi_3$, this is $({}^{+2.8}_{-1.3})^\circ$. This additional uncertainty is added in quadrature to the result in Table~\ref{tab:DhDstK1DResult1}, giving the final result of $\phi_3 = (75.2\pm 7.6)^\circ$.
The additional uncertainties for other parameters are negligible compared to the nominal uncertainties, as shown in Table~\ref{tab:CorrTest}. We ignore the correlation of systematic uncertainties between different measurements from the same experiments. As the precision of these measurements is dominated by statistical uncertainties, the impact of neglecting such correlations is negligible.

\begin{table}[htp]
    \centering
    \caption{Additional uncertainties on all the parameters due to unknown correlations.}
    \vspace{0.5cm}
    \begin{tabular}{c c c c c c c c }
    \hline
    Parameters & $\phi_3 (^\circ)$ & $r_B^{DK}$  & $\delta_B^{DK} (^\circ)$  & $r_B^{D\pi}$  & $\delta_B^{D\pi} (^\circ)$ & $r_B^{D^*K}$ & $\delta_B^{D^*K} (^\circ)$ \\
    \hline
    %Uncertainty & ${}^{+2.4}_{-1.1}$ & ${}^{+0.006}_{-0.004}$ &${}^{+1.5}_{-0.6}$ &${}^{+0.0003}_{-0.0006}$ &${}^{+0.0}_{-0.5}$ &${}^{+0.003}_{-0.000}$ &${}^{+1.0}_{-0.0}$ \\
    Uncertainty& ${}^{+2.8}_{-1.3}$ & ${}^{+0.005}_{-0.005}$& ${}^{+1.7}_{-0.7}$& ${}^{+0.0001}_{-0.001}$& ${}^{+0}_{-0.7}$& ${}^{+0.002}_{-0}$& ${}^{+0.2}_{-0.1}$\\%updated with mixing
    \hline
    \end{tabular}
    \label{tab:CorrTest}
\end{table}

%\clearpage

\subsection{Discussion} 
\label{sec:discussion}

Our combined determination of $\phi_3$ alone is consistent with the current world average value within two standard deviations ($\sigma$)~\cite{hflav}. However, the collective agreement of our full set of results with the world average values is poor, with a $p$-value of 0.45\%.\footnote{In the comparison, we do not include the correlation due to the common Belle results used in our combination and in the current world average.} In addition, the sensitivity obtained for $\phi_3$ is better than originally anticipated~\cite{snowmass}.

The discrepancy with the world average arises primarily from the large values obtained for $r_B^{D\pi}$ and $\delta_B^{D\pi}$, which deviate by $2.2\sigma$ and $4.0\sigma$, respectively, from the world average values.\footnote{In this context, the ``world average" refers to the LHCb average of \(r_B^{D\pi}\) and \(\delta_B^{D\pi}\), as there are currently no world average values for \(r_B^{D\pi}\) and \(\delta_B^{D\pi}\)~\cite{hflav}.} 
These parameters are correlated: a large $r_B^{D\pi}$ value results in a smaller uncertainty on $\delta_B^{D\pi}$, thus giving a larger deviation for this parameter.
%Surprisingly, we achieve a higher sensitivity for $\phi_3$ than initially anticipated, considering the WA values of all parameters and the uncertainties of the input parameters. This enhanced sensitivity arises because our fit favours a larger $r_B^{DK}$ value by $1.5\sigma$ compared to the WA. Additionally, both $r_B^{D\pi}$ and $\delta_B^{D\pi}$ exhibit values that are larger than the WA values, with deviations of $2.2\sigma$ and $4.0\sigma$, respectively, further enhancing the precision of our determination.  
The large $r_B^{D\pi}$ value is mainly determined by the values of input observables in Equation~\ref{eq:xi_param}. While these observables agree within $2\sigma$ with the values reported by the only other measurement available with this parametrization~\cite{lhcb2021}, their large central values lead to an unexpectedly high $r_B^{D\pi}$ value. For instance, the expected value for the $r_B^{D\pi}/r_B^{DK}$ ratio is approximately 1/20, considering only the ratio of CKM matrix-elements involved in the amplitudes, i.e., $\vert V^{*}_{ub} V_{cd}/V^{*}_{cb} V_{ud} \vert$. However, we obtain approximately 1/7 for this ratio in our combination.  We demonstrate that the departure of our results from the world averages is due to our higher measured value of $r_B^{D\pi}$ by repeating the combination after constraining $r_B^{D\pi}$ to its expected value $r_B^{D\pi} = 0.0053 \pm 0.0007$, estimated using the known branching fractions of various $B \to D K$ and $B \to D \pi$ decays and SU(3) symmetry~\cite{rBDpi}. With this additional constraint, our combined $\phi_3$ value is $(78.7 \pm 8.1)^\circ$, which is not significantly different from the nominal result. However, the resulting increased uncertainty on $\delta_B^{D\pi}$ yields better agreement of the full set of results with the world averages, with a $p$-value of 13\%.

The second aspect that requires further checks is our better-than-expected sensitivity to $\phi_3$~\cite{snowmass}. We investigate this by studying separately the contribution of the individual inputs to the $\phi_3$ precision, as shown in Table~\ref{tab:multicombiner}. The precision on $\phi_3$ improves significantly, from $11^\circ$ to $8.3^\circ$, when the ADS inputs from $B^+ \to D(\to K^+\pi^-)h^+$ are combined with the BPGGSZ inputs. This enhancement is driven by the  $R_{\textrm{ADS}}$ observable of the $B^+ \to D\pi^+$ channel. The relation of this observable with hadronic parameters in the absence of $D^0-\overline{D}^0$ mixing is
\begin{equation}
    R_{\rm ADS}^{D\pi, K\pi} = (r^{D\pi}_{B})^{2} + (r^{K\pi}_{D})^{2} + 2r^{D\pi}_{B}r^{K\pi}_{D}\cos(\delta^{D\pi}_{B}+\delta^{K\pi}_{D})\cos\phi_{3}.
\end{equation}
Substitution in the above equation of our $r_B^{D\pi}$ value, the auxiliary input $r_D^{K\pi}$, and their uncertainties, greatly enhances the precision on the interference (last) term as compared to the $B^+ \to DK^+$ case. Furthermore, our value $\delta_B^{D\pi} \approx 347^\circ$ leads to a precise determination of $\cos(\delta_B^{D\pi}+\delta_D^{K\pi})$ close to one. The combined effect of both factors improves the precision of our $\phi_3$ result. The ADS contribution to the sensitivity of $\phi_3$ is primarily attributed to three elements: the small relative uncertainty of the large $r_B^{D\pi}$ value favoured by Belle and Belle II measurements, the availability of a precise value of $r_D^{K\pi}$ from global averages, and the large $\delta_B^{D\pi}$ value favoured by Belle and Belle~II measurements.
%\comment{These reasons also lead to the non-negligible impact of $D^0-\bar{D}^0$ mixing on $\phi_3$ from this single observable $R_{\rm ADS}^{D\pi, K\pi}$.}
%It is worth noting that in the future, as larger data sets become available, the values of the hadronic parameters $r_B^{D\pi}$ and $\delta_B^{D\pi}$ may converge towards the WA values. In such a scenario, the contribution of ADS($K\pi$) to the precision of $\phi_3$ would be smaller.

Finally, we check the impact of our large $r_B^{DK}$ value, which is $1.5\sigma$ higher than the world average. We generate simulated experiments assuming world average values for $\phi_3$ and of the hadronic parameters. %This leads to a precision of $12^\circ$. 
Repeating the analysis on these gives $12^\circ$ precision on $\phi_3$.
We repeat the study assuming our combined value for $r_B^{DK}$ and observe that the precision improves to $10.1^\circ$. Finally, we repeat the study by assuming our combined values for $\phi_3, r_{B}^{DK}, \delta_B^{DK}, r_{B}^{D\pi},$ and $\delta_B^{D\pi}$ and obtain a precision of $7.1^\circ$, which is consistent with our nominal results. 
%the numbers are unchanged.

By employing simulated experiments generated using world average values, we extrapolate this result to future sample sizes, and find the current uncertainty on $\phi_3$ to be consistent with recent projections~\cite{snowmass}.

\subsection{Coverage study}
\label{sec:coverage}
We check the statistical coverage of the fit by generating simulated experiments at the best-fit point.
%Sometimes, the statistical coverage of the used methods cannot be trusted. Hence, we perform the coverage study computed at the best-fit point for this combination. 
%We generate simulated experiments and compute the $p$-value of the best-fit point. 
The coverage is then defined as the fraction of times the \(1\sigma\) or \(2\sigma\) interval of the fitted \(\phi_3\) contains the true value of \(\phi_3\).
%the best-fit value of $\phi_3$ has a larger 1$-$CL than $68.3\%$ and \comment{$ 95.4\%$}, for 1$\sigma$ and 2$\sigma$, respectively. 
We perform the coverage study assuming that the true values of all relevant parameters are the values measured in our data. The resulting fractions are $0.672 \pm 0.004$ and $0.951 \pm 0.002$ for 1$\sigma$ and 2$\sigma$ coverage, respectively.
%The results of the coverage study done assuming that the true values of all relevant parameters are the values measured in our data are shown in Table~\ref{tab:coveragestudy}. 
%The \textsc{Prob} method exhibits a slight under-coverage, but its impact on our provided confidence intervals is negligible.

\begin{figure}[htp]
\centering
\begin{tabular}{c c}
    \begin{overpic}[width=0.45\textwidth]{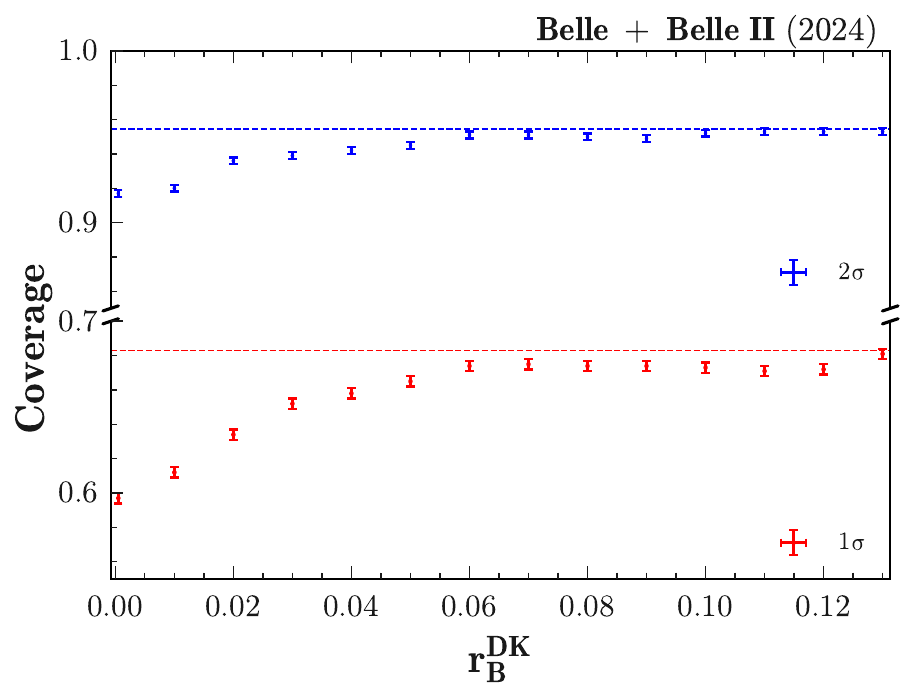}\end{overpic}
    \begin{overpic}[width=0.45\textwidth]{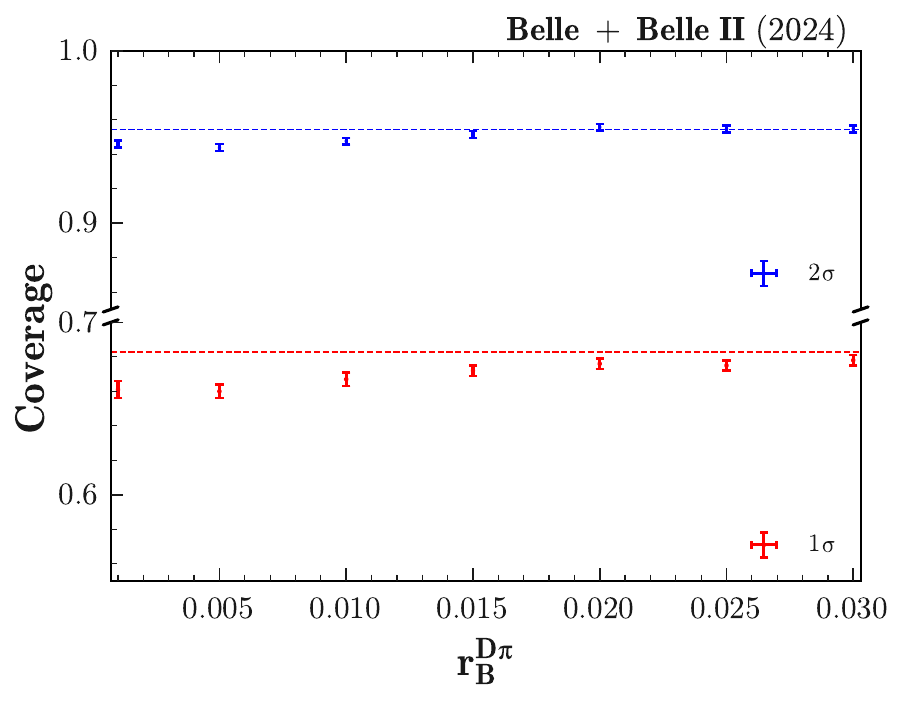}\end{overpic}
\end{tabular}
 \caption{Dependence of the coverage of the $\phi_3$ interval on $r_B^{DK}$ (left) and $r_B^{D\pi}$ (right). The dashed horizontal lines show the nominal coverage at 1 (red) and 2$\sigma$ (blue) for confidence levels of 68.3\% and 95.4\%, respectively. The y-axis range is split to make the error bars visible.}
  \label{fig:coverage}
\end{figure}

We also test coverage at $r_B^{DK, D\pi}$ values other than the best-fit point; the 1$\sigma$ and 2$\sigma$ coverage is shown in Figure~\ref{fig:coverage}.
The coverage for the combination degrades as the true values of $r_B^{DK, D\pi}$ become smaller. This behaviour has previously been observed by the CKMfitter group~\cite{ckmfitter} and the LHCb experiment~\cite{Aaij_2016}. The fitted
values found in this combination, $r_B^{DK} = 0.115$ and $r_B^{D\pi} = 0.016$, are well within the regime of accurate coverage. No correction for under-coverage is applied to the
confidence intervals quoted in Tables~\ref{tab:DhDstK1DResult1} and~\ref{tab:multicombiner}.
%In addition, we also assess coverage by randomly sampling true values for $r_B^{DK}$, $r_B^{D\pi}$, and $\delta_B^{D\pi}$, as these parameters exhibit significant differences compared to the world average values. No obvious undercoverage is found.
%Again, we observe slight under-coverage with the \textsc{Prob} method, resembling the behaviour shown in Figure~\ref{fig:coverage}, yet its impact is negligible.

%%%%%%%%%%%%%%%%%%%%%%%%%%%%%%%%%%%%%%%%%%%%%%%%%%%%%%%%%%%%%%%%%%%%%%%%
\section{Summary}
\label{sec:summary}
In summary, we report the value of the CKM angle $\phi_3$ by combining existing Belle and Belle~II measurements and auxiliary $D$-decay information from other experiments. This combination includes inputs from a number of $B^+ \to Dh^+$ and $B^+ \to D^*K^+$ decay modes. The resulting value for $\phi_3$ is $(75.2\pm 7.6)^\circ$, which is consistent with the world average value within 1.1$\sigma$~\cite{hflav}. We obtain higher values of the parameters $r_B^{DK}$, $r_B^{D\pi}$, and $\delta_B^{D\pi}$ compared to the world average values, with deviations of $1.5\sigma$, $2.2\sigma$, and $4.0\sigma$, respectively. We demonstrate that these differences are likely to be due to our large measured value of $r_B^{D\pi}$. In addition, we achieve a better than anticipated precision on $\phi_3$. This is attributed to the combined effect of a precise $r_D^{K\pi}$ value and larger than expected values of $r_B^{D\pi}$ and $\delta_B^{D\pi}$, which have a significant impact on the ADS ($K^+\pi^-$) final state.

\clearpage

\acknowledgments
% Policy from October 20, 2022
This work, based on data collected using the Belle II detector, which was built and commissioned prior to March 2019,
and data collected using the Belle detector, which was operated until June 2010,
was supported by
%Armenia
Higher Education and Science Committee of the Republic of Armenia Grant No.~23LCG-1C011;
%Australia
Australian Research Council and Research Grants
No.~DP200101792, % Jackson
No.~DP210101900, % Urquijo
No.~DP210102831, % Sevior
No.~DE220100462, % Hsu
No.~LE210100098, % Infrastructure
and
No.~LE230100085; % Infrastructure
%Austria
Austrian Federal Ministry of Education, Science and Research,
Austrian Science Fund
No.~P~31361-N36
and
No.~J4625-N,
and
Horizon 2020 ERC Starting Grant No.~947006 ``InterLeptons'';
%Canada
Natural Sciences and Engineering Research Council of Canada, Compute Canada and CANARIE;
%China
National Key R\&D Program of China under Contract No.~2022YFA1601903,
National Natural Science Foundation of China and Research Grants
No.~11575017,
No.~11761141009,
No.~11705209,
No.~11975076,
No.~12135005,
No.~12150004,
No.~12161141008,
and
No.~12175041,
and Shandong Provincial Natural Science Foundation Project~ZR2022JQ02;
%Czech Republic
the Czech Science Foundation Grant No.~22-18469S;
%EU
European Research Council, Seventh Framework PIEF-GA-2013-622527,
Horizon 2020 ERC-Advanced Grants No.~267104 and No.~884719,
Horizon 2020 ERC-Consolidator Grant No.~819127,
Horizon 2020 Marie Sklodowska-Curie Grant Agreement No.~700525 ``NIOBE''
and
No.~101026516,
and
Horizon 2020 Marie Sklodowska-Curie RISE project JENNIFER2 Grant Agreement No.~822070 (European grants);
%France
L'Institut National de Physique Nucl\'{e}aire et de Physique des Particules (IN2P3) du CNRS
and
L'Agence Nationale de la Recherche (ANR) under grant ANR-21-CE31-0009 (France);
%Germany
BMBF, DFG, HGF, MPG, and AvH Foundation (Germany);
%India
Department of Atomic Energy under Project Identification No.~RTI 4002,
Department of Science and Technology,
and
UPES SEED funding programs
No.~UPES/R\&D-SEED-INFRA/17052023/01 and
No.~UPES/R\&D-SOE/20062022/06 (India);
%Israel
Israel Science Foundation Grant No.~2476/17,
U.S.-Israel Binational Science Foundation Grant No.~2016113, and
Israel Ministry of Science Grant No.~3-16543;
%Italy
Istituto Nazionale di Fisica Nucleare and the Research Grants BELLE2;
%Japan
Japan Society for the Promotion of Science, Grant-in-Aid for Scientific Research Grants
No.~16H03968,
No.~16H03993,
No.~16H06492,
No.~16K05323,
No.~17H01133,
No.~17H05405,
No.~18K03621,
No.~18H03710,
No.~18H05226,
No.~19H00682, % Niigata
No.~20H05850,
No.~20H05858,
No.~22H00144,
No.~22K14056,
No.~22K21347,
No.~23H05433,
No.~26220706,
and
No.~26400255,
%the National Institute of Informatics, and Science Information NETwork 5 (SINET5), 
and
the Ministry of Education, Culture, Sports, Science, and Technology (MEXT) of Japan;  
%Korea
National Research Foundation (NRF) of Korea Grants
No.~2016R1\-D1A1B\-02012900,
No.~2018R1\-A2B\-3003643,
No.~2018R1\-A6A1A\-06024970,
No.~2019R1\-I1A3A\-01058933,
No.~2021R1\-A6A1A\-03043957,
No.~2021R1\-F1A\-1060423,
No.~2021R1\-F1A\-1064008,
No.~2022R1\-A2C\-1003993,
and
No.~RS-2022-00197659,
Radiation Science Research Institute,
Foreign Large-Size Research Facility Application Supporting project,
the Global Science Experimental Data Hub Center of the Korea Institute of Science and Technology Information
and
KREONET/GLORIAD;
%Malaysia
Universiti Malaya RU grant, Akademi Sains Malaysia, and Ministry of Education Malaysia;
%Mexico
% CINVESTAV-IPN, UNAM, UAS, BUAP and CONACYT are funded under
Frontiers of Science Program Contracts
No.~FOINS-296,
No.~CB-221329,
No.~CB-236394,
No.~CB-254409,
and
No.~CB-180023, and SEP-CINVESTAV Research Grant No.~237 (Mexico);
%Poland
the Polish Ministry of Science and Higher Education and the National Science Center;
%Russia
the Ministry of Science and Higher Education of the Russian Federation
and
the HSE University Basic Research Program, Moscow;
%Saudi Arabia
University of Tabuk Research Grants
No.~S-0256-1438 and No.~S-0280-1439 (Saudi Arabia);
%Slovenia
Slovenian Research Agency and Research Grants
No.~J1-9124
and
No.~P1-0135;
%Spain
Ikerbasque, Basque Foundation for Science,
the State Agency for Research of the Spanish Ministry of Science and Innovation through Grant No. PID2022-136510NB-C33,
Agencia Estatal de Investigacion, Spain
Grant No.~RYC2020-029875-I
and
Generalitat Valenciana, Spain
Grant No.~CIDEGENT/2018/020;
%Swiss (Belle 1)
the Swiss National Science Foundation;
%Taiwan
National Science and Technology Council,
and
Ministry of Education (Taiwan);
%Thailand
Thailand Center of Excellence in Physics;
%Turkey
TUBITAK ULAKBIM (Turkey);
%Ukraine
National Research Foundation of Ukraine, Project No.~2020.02/0257,
and
Ministry of Education and Science of Ukraine;
%USA
the U.S. National Science Foundation and Research Grants
No.~PHY-1913789 % Indiana CEEM
and
No.~PHY-2111604, % Luther
and the U.S. Department of Energy and Research Awards
No.~DE-AC06-76RLO1830, % PNNL
No.~DE-SC0007983, % Wayne State
No.~DE-SC0009824, % Florida
No.~DE-SC0009973, % VPI
No.~DE-SC0010007, % Duke
No.~DE-SC0010073, % South Carolina
No.~DE-SC0010118, % Carnegie Mellon
No.~DE-SC0010504, % Hawaii
No.~DE-SC0011784, % Cincinnati
No.~DE-SC0012704, % BNL
No.~DE-SC0019230, % Duke
No.~DE-SC0021274, % Mississippi
No.~DE-SC0021616, % Mississippi
No.~DE-SC0022350, % Louisville
No.~DE-SC0023470; % South Alabama
%last group
and
%Vietnam
the Vietnam Academy of Science and Technology (VAST) under Grants
No.~NVCC.05.12/22-23
and
No.~DL0000.02/24-25.

% Policy from October 20, 2022
These acknowledgements are not to be interpreted as an endorsement of any statement made
by any of our institutes, funding agencies, governments, or their representatives.

We thank the SuperKEKB team for delivering high-luminosity collisions;
the KEK cryogenics group for the efficient operation of the detector solenoid magnet;
the KEK Computer Research Center for on-site computing support; the NII for SINET6 network support;
and the raw-data centers hosted by BNL, DESY, GridKa, IN2P3, INFN, 
PNNL/EMSL, 
and the University of Victoria.

%\paragraph{Note added.} This is also a good position for notes added after the paper has been written.

\clearpage
\bibliographystyle{JHEP}
\bibliography{references}
%%%%%%%%%%%%%%%%%%%%%%%%%%%%%%%%%%%%%%%%%%%%%%%%%%%%%%%%%%%%%%%%%%%%%%%%%%
\clearpage
\appendix
\notocsection{$\chi^2$ of each input measurement}
\label{app:chisq}
We show the $\chi^2$ value for each input measurement in Table.~\ref{tab:Chisq}.

\begin{table}[!h]
    \centering
    \caption{Auxiliary input observables and their values used in the $\phi_3$ combination.}
    \vspace{0.5cm}
    \begin{tabular}{c| c c c}
    \hline
     & Measurement  & $\chi^2$ & No. of obs. \\
     \hline
     \multirow{8}{*}{\rotatebox[origin=c]{90}{Belle and Belle II}}& $B^+ \to Dh^+, D \to \KS \pi^0, K^-K^+$ & 7.50 & 4\\
     & $B^+ \to Dh^+, D \to K^+\pi^-$ & 1.33 & 4\\
     & $B^+ \to Dh^+, D \to K^+\pi^-\pi^0$ & 5.81 & 4\\
     & $B^+ \to Dh^+, D \to \KS K^-\pi^+$ & 7.52 & 7 \\
     & $B^+ \to Dh^+ D \to \KS h^-h^+$ & 2.07 & 6\\
     & $B^+ \to Dh^+, D \to \KS \pi^-\pi^+\pi^0$ & 9.53 & 8\\
     & $B^+ \to D^* K^+, \Dstar\to D\piz, D \to \KS\piz, \KS\phi, \KS\omega, K^-K^+, \pi^-\pi^+$  & 1.45 & 4 \\
     & $B^+ \to D^* K^+, \Dstar\to D\piz, D\gamma, D \to \KS \pi^-\pi^+$ & 1.42 & 8\\
    \hline
    \multirow{5}{*}{\rotatebox[origin=c]{90}{External}}& $D\to K^+\pi^-$ & 0.03 & 2 \\
    & $D\to K^+\pi^-\pi^0$ & 0.14 & 3 \\
    & $D-\overline{D}$ mixing & 0.02 & 4 \\
    & $D\to \KS K^-\pi^+$ & 1.17 & 4 \\
    & $R_{\textrm{GLS}}$ & 0.12 & 1 \\
    \hline
    & Total & 38.1 &  59 \\
    \hline
    \end{tabular}
    \label{tab:Chisq}
\end{table}

\begin{figure}[htp]
  \centering
  \begin{overpic}[width=0.25\textwidth]{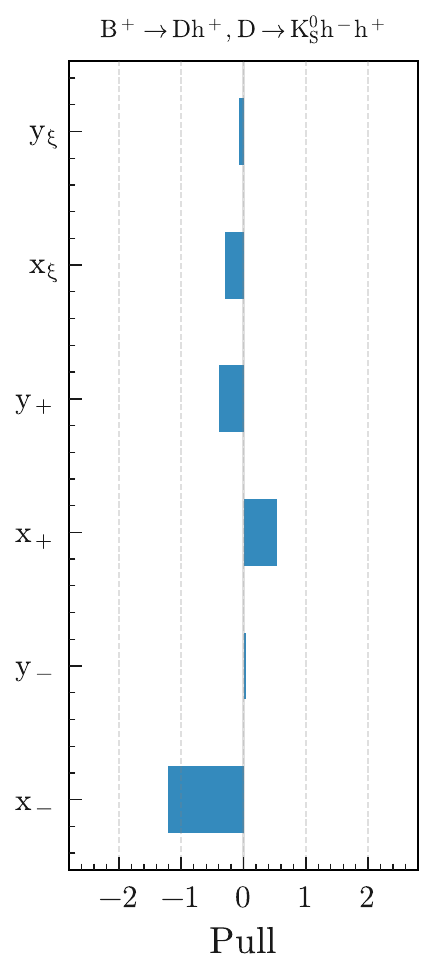}\put(75,45){}\end{overpic}
  \begin{overpic}[width=0.25\textwidth]{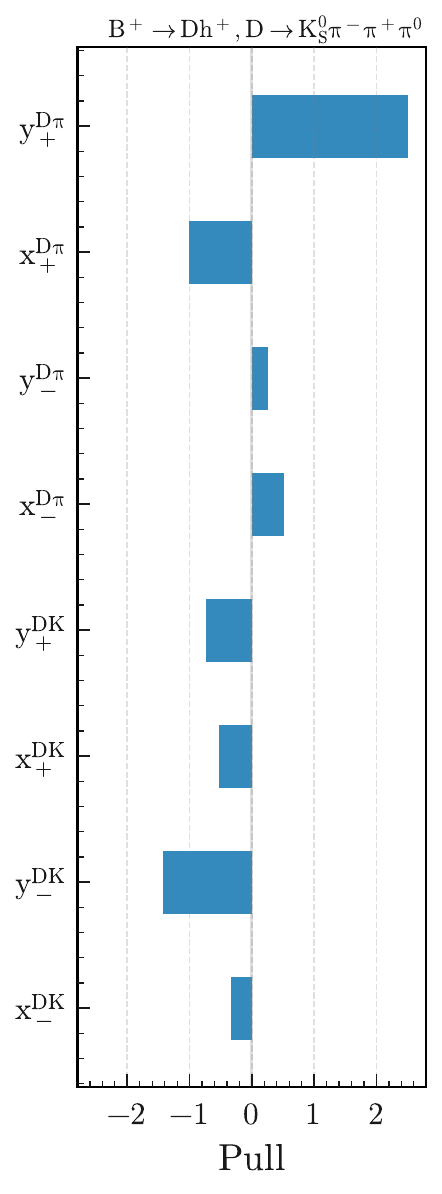}\put(75,45){}\end{overpic}
  \begin{overpic}[width=0.25\textwidth]{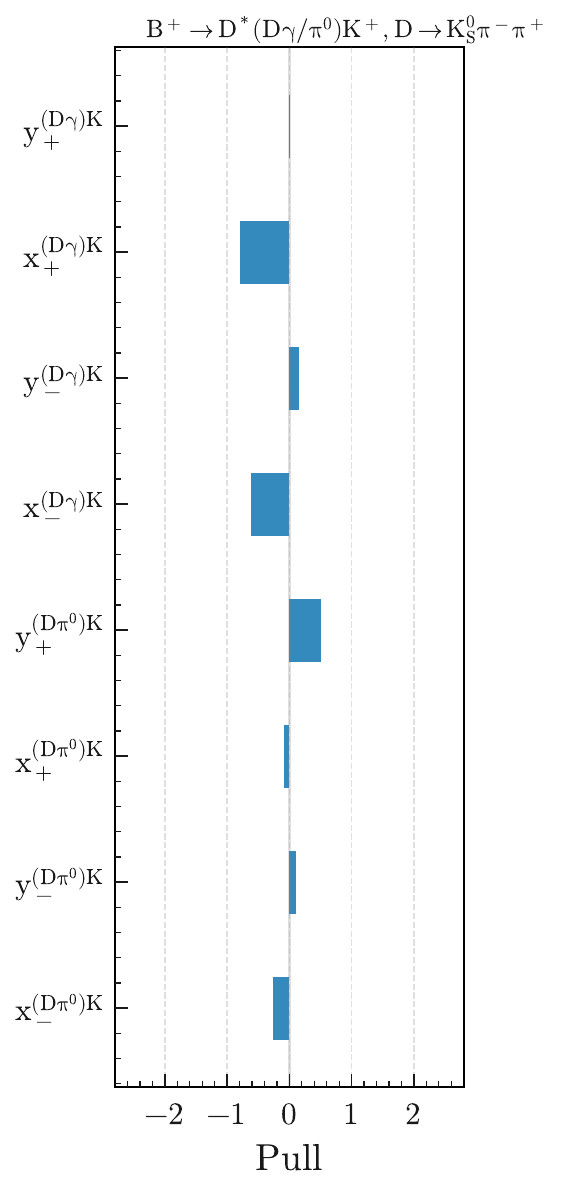}\put(75,45){}\end{overpic}
  \begin{overpic}[width=0.25\textwidth]{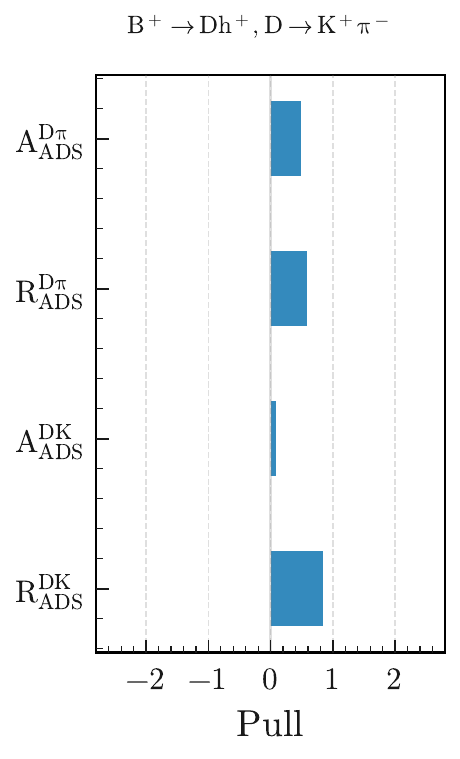}\put(75,45){}\end{overpic}
  \begin{overpic}[width=0.25\textwidth]{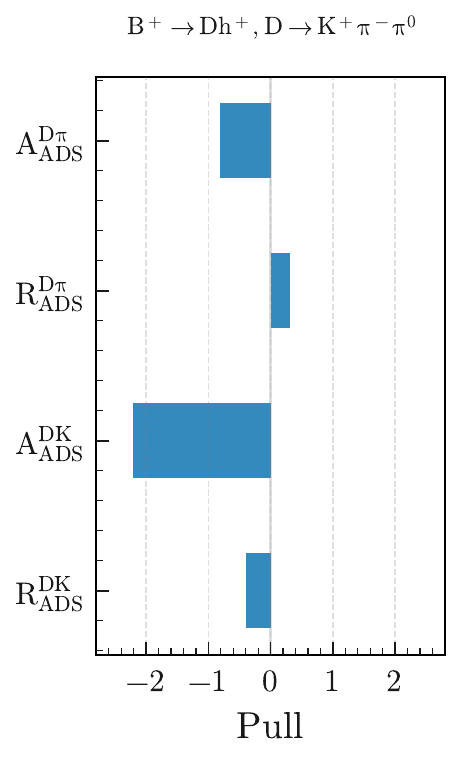}\put(75,45){}\end{overpic}
  \begin{overpic}[width=0.25\textwidth]{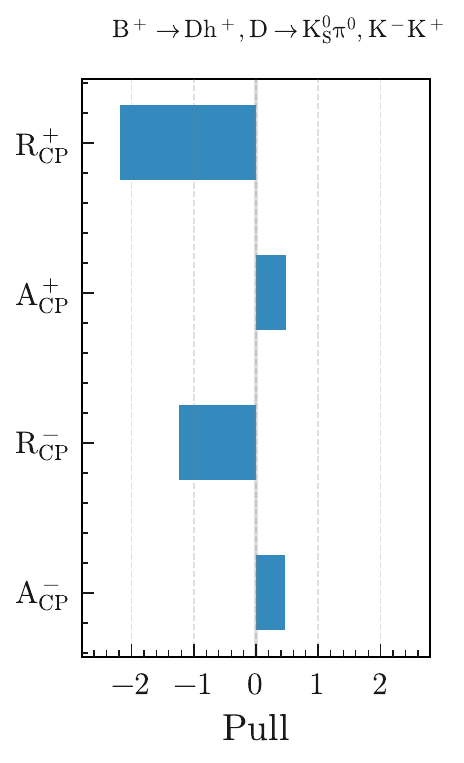}\put(75,45){}\end{overpic}
  \begin{overpic}[width=0.25\textwidth]{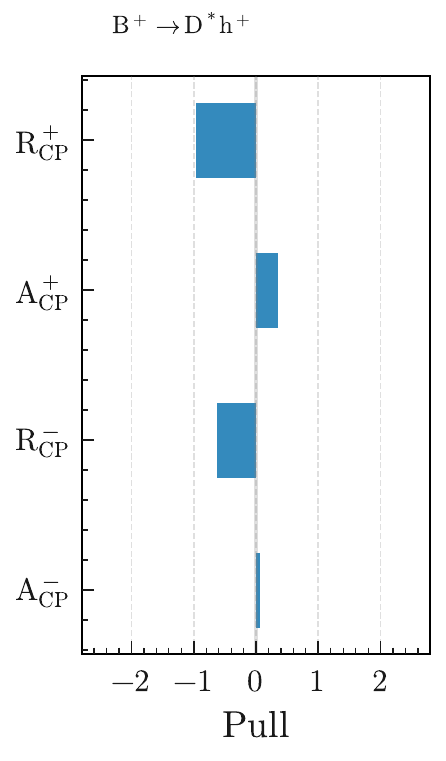}\put(75,45){}\end{overpic}
  \begin{overpic}[width=0.25\textwidth]{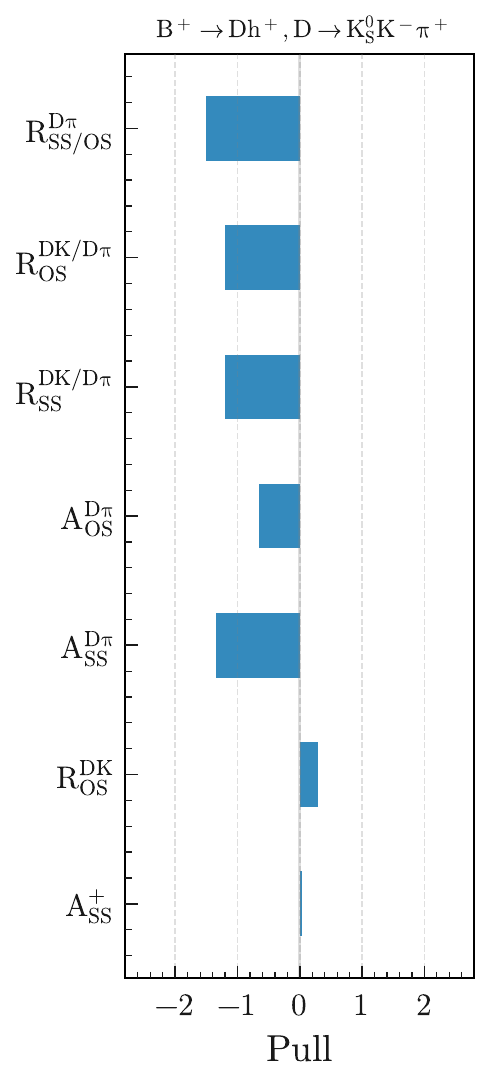}\put(75,45){}\end{overpic}
  \caption{Pulls of the input observables from Belle and Belle II.}
  \label{fig:Pulls}
\end{figure}

\notocsection{Pull distribution of each input observable}
\label{app:pull}
We show the pull of each input observable with respect to the global best-fit point in Figure~\ref{fig:Pulls}. 
%The pull is defined as $(A_{\textrm{obs}} - A_{\textrm{fit}})/\sigma({\textrm{obs}})$, where $A_{\textrm{obs}}$ and $A_{\textrm{fit}}$ are the input value and the best-fit value, respectively, and $\sigma({\textrm{obs}})$ is the measurement uncertainty.

\clearpage
\notocsection{Relationships among parameters and observables}
\label{app:equations}
We list the equations that give the relationships among the parameters of interest and observables in the $B$ decay channels. For simplicity, the equations are given in the absence of $D^0-\overline{D}^0$ mixing. In order to include the small effects from $D^0-\overline{D}^0$ mixing, the equations should be modified following the recommendation in Ref.~\cite{mixing}.

\begin{itemize}
    \item $B^+ \to Dh^+, D\to\KS h^-h^+$ observables
\begin{align}\label{eq:bpggsz}
\begin{split}
x^{DK}_{\pm} & = r^{DK}_{B}\cos(\delta^{DK}_{B} \pm \phi_{3}),\\
y^{DK}_{\pm} & = r^{DK}_{B}\sin(\delta^{DK}_{B} \pm \phi_{3}),\\
x_\xi^{D\pi} & = (r^{D\pi}_{B}/r^{DK}_{B})\cos(\delta^{D\pi}_{B} - \delta^{DK}_{B}),\\
y_\xi^{D\pi} & = (r^{D\pi}_{B}/r^{DK}_{B})\sin(\delta^{D\pi}_{B} - \delta^{DK}_{B}).\\
\end{split}
\end{align}
    \item $B^+\to Dh^+, D\to\KS\pi^-\pi^+\piz$ observables
\begin{align}\label{eq:bpggsz2}
\begin{split}
x^{DK}_{\pm} & = r^{DK}_{B}\cos(\delta^{DK}_{B} \pm \phi_{3}),\\
y^{DK}_{\pm} & = r^{DK}_{B}\sin(\delta^{DK}_{B} \pm \phi_{3}),\\
x^{D\pi}_{\pm} & = r^{D\pi}_{B}\cos(\delta^{D\pi}_{B} \pm \phi_{3}),\\
y^{D\pi}_{\pm} & = r^{D\pi}_{B}\sin(\delta^{D\pi}_{B} \pm \phi_{3}).
\end{split}
\end{align}
    \item $B^+\to D^*K^+, D\to\KS\pi^-\pi^+$ observables
\begin{align}\label{eq:anton}
\begin{split}
x^{(D\piz)K}_{\pm} & = r^{D^*K}_{B}\cos(\delta^{D^*K}_{B} \pm \phi_{3}),\\
y^{(D\piz)K}_{\pm} & = r^{D^*K}_{B}\sin(\delta^{D^*K}_{B} \pm \phi_{3}),\\
x^{(D\gamma)K}_{\pm} & = -r^{D^*K}_{B}\cos(\delta^{D^*K}_{B} \pm \phi_{3}),\\
y^{(D\gamma)K}_{\pm} & = -r^{D^*K}_{B}\sin(\delta^{D^*K}_{B} \pm \phi_{3}),\\
\end{split}
\end{align}
    \item $B^+\to Dh^+, D\to K^+\pi^-$ observables
\begin{align}\label{eq:ads1}
\begin{split}
R_{\rm ADS}^{DK, K\pi} &= (r^{DK}_{B})^{2} + (r^{K\pi}_{D})^{2} + 2r^{DK}_{B}r^{K\pi}_{D}\cos(\delta^{DK}_{B}+\delta^{K\pi}_{D})\cos\phi_{3},\\
A_{\rm ADS}^{DK, K\pi} &= 2r^{DK}_{B}r^{K\pi}_{D}\sin(\delta^{DK}_{B}+\delta^{K\pi}_{D})\sin\phi_{3}/R_{\rm ADS}^{DK, K\pi},\\
R_{\rm ADS}^{D\pi, K\pi} &= (r^{D\pi}_{B})^{2} + (r^{K\pi}_{D})^{2} + 2r^{D\pi}_{B}r^{K\pi}_{D}\cos(\delta^{D\pi}_{B}+\delta^{K\pi}_{D})\cos\phi_{3},\\
A_{\rm ADS}^{D\pi, K\pi} &= 2r^{D\pi}_{B}r^{K\pi}_{D}\sin(\delta^{D\pi}_{B}+\delta^{K\pi}_{D})\sin\phi_{3}/R_{\rm ADS}^{D\pi, K\pi}.
\end{split}
\end{align}
    \item $B^+\to Dh^+, D\to K^+\pi^-\piz$ observables
\begin{align}\label{eq:ads2}
\begin{split}
R_{\rm ADS}^{DK, K\pi\piz} &= (r^{DK}_{B})^{2} + (r^{K\pi\piz}_{D})^{2} + 2r^{DK}_{B}r^{K\pi\piz}_{D}\kappa^{K\pi\piz}_{D}\cos(\delta^{DK}_{B}+\delta^{K\pi\piz}_{D})\cos\phi_{3},\\
A_{\rm ADS}^{DK, K\pi\piz} &= 2r^{DK}_{B}r^{K\pi\piz}_{D}\kappa^{K\pi\piz}_{D}\sin(\delta^{DK}_{B}+\delta^{K\pi\piz}_{D})\sin\phi_{3}/R_{\rm ADS}^{DK, K\pi},\\
R_{\rm ADS}^{D\pi, K\pi\piz} &= (r^{D\pi}_{B})^{2} + (r^{K\pi\piz}_{D})^{2} + 2r^{D\pi}_{B}r^{K\pi}_{D}\kappa^{K\pi\piz}_{D}\cos(\delta^{D\pi\piz}_{B}+\delta^{K\pi\piz}_{D})\cos\phi_{3},\\
A_{\rm ADS}^{D\pi, K\pi\piz} &= 2r^{D\pi}_{B}r^{K\pi\piz}_{D}\kappa^{K\pi\piz}_{D}\sin(\delta^{D\pi\piz}_{B}+\delta^{K\pi\piz}_{D})\sin\phi_{3}/R_{\rm ADS}^{D\pi, K\pi\piz}.
\end{split}
\end{align}
    \item $B^+\to Dh^+, D\to \KS\piz, K^-K^+$ observables
\begin{align}\label{eq:glw1}
\begin{split}
\mathcal{R}_{\CP\pm}&= 1 + (r^{DK}_{B})^{2} + 2 \eta_{\CP} r^{DK}_{B} \cos(\delta^{DK}_{B}) \cos \phi_{3},\\
A_{\CP\pm}&= 2 \eta_{\CP} r^{DK}_{B} \sin(\delta^{DK}_{B}) \sin \phi_{3}/ R_{\CP\pm},
\end{split}
\end{align}
where $\eta_{\CP}$ denotes the $\CP$ eigenvalue of the $D$ decay.
    \item $B^+\to D^*h^+, D\to \KS\piz, K^-K^+, \KS\phi, \KS\omega, \pi^-\pi^+$ observables
\begin{align}\label{eq:glw2}
\begin{split}
R_{\CP\pm}&= 1 + (r^{D^*K}_{B})^{2} + 2 \eta_{\CP} r^{D^*K}_{B} \cos(\delta^{D^*K}_{B}) \cos \phi_{3},\\
A_{\CP\pm}&= 2 \eta_{\CP} r^{D^*K}_{B} \sin(\delta^{D^*K}_{B}) \sin \phi_{3}/ R_{\CP\pm}.
\end{split}
\end{align}
    \item $B^+\to Dh^+, D\to \KS K^-\pi^+$ observables
\begin{align}\label{eq:gls}
\begin{split}
A^{DK}_{SS} & = \frac{2r^{DK}_Br^{\KS K\pi}_D\kappa^{\KS K\pi}_D\sin(\delta^{DK}_B-\delta^{\KS K\pi}_D)\sin\phi_3}{1+(r^{DK}_B)^2(r^{\KS K\pi}_D)^2+2r^{DK}_Br^{\KS K\pi}_D\kappa^{\KS K\pi}_D\cos(\delta^{DK}_B-\delta^{\KS K\pi}_D)\cos\phi_3},  \\
A^{DK}_{OS} & = \frac{2r^{DK}_Br^{\KS K\pi}_D\kappa^{\KS K\pi}_D\sin(\delta^{DK}_B+\delta^{\KS K\pi}_D)\sin\phi_3}{(r^{DK}_B)^2+(r^{\KS K\pi}_D)^2+2r^{DK}_Br^{\KS K\pi}_D\kappa^{\KS K\pi}_D\cos(\delta^{DK}_B+\delta^{\KS K\pi}_D)\cos\phi_3}, \\ 
A^{D\pi}_{SS} & = \frac{2r^{D\pi}_Br^{\KS K\pi}_D\kappa^{\KS K\pi}_D\sin(\delta^{D\pi}_B-\delta^{\KS K\pi}_D)\sin\phi_3}{1+(r^{D\pi}_B)^2(r^{\KS K\pi}_D)^2+2r^{D\pi}_Br^{\KS K\pi}_D\kappa^{\KS K\pi}_D\cos(\delta^{D\pi}_B-\delta^{\KS K\pi}_D)\cos\phi_3},  \\
A^{D\pi}_{OS} & = \frac{2r^{D\pi}_Br^{\KS K\pi}_D\kappa^{\KS K\pi}_D\sin(\delta^{D\pi}_B+\delta^{\KS K\pi}_D)\sin\phi_3}{(r^{D\pi}_B)^2+(r^{\KS K\pi}_D)^2+2r^{D\pi}_Br^{\KS K\pi}_D\kappa^{\KS K\pi}_D\cos(\delta^{D\pi}_B+\delta^{\KS K\pi}_D)\cos\phi_3}. \\ 
R^{DK/D\pi}_{SS} & =R_{\textrm{GLS}}\frac{1+(r^{DK}_B)^2(r^{\KS K\pi}_D)^2+2r^{DK}_Br^{\KS K\pi}_D\kappa^{\KS K\pi}_D\cos(\delta^{DK}_B-\delta^{\KS K\pi}_D)\cos\phi_3}{1+(r^{D\pi}_B)^2(r^{\KS K\pi}_D)^2+2r^{D\pi}_Br^{\KS K\pi}_D\kappa^{\KS K\pi}_D\cos(\delta^{D\pi}_B-\delta^{\KS K\pi}_D)\cos\phi_3}, \\
R^{DK/D\pi}_{OS} & =R_{\textrm{GLS}}\frac{(r^{DK}_B)^2+(r^{\KS K\pi}_D)^2+2r^{DK}_Br^{\KS K\pi}_D\kappa^{\KS K\pi}_D\cos(\delta^{DK}_B+\delta^{\KS K\pi}_D)\cos\phi_3}{(r^{D\pi}_B)^2+(r^{\KS K\pi}_D)^2+2r^{D\pi}_Br^{\KS K\pi}_D\kappa^{\KS K\pi}_D\cos(\delta^{D\pi}_B+\delta^{\KS K\pi}_D)\cos\phi_3}, \\
R^{D\pi}_{SS/OS} & =\frac{1+(r^{D\pi}_B)^2(r^{\KS K\pi}_D)^2+2r^{D\pi}_Br^{\KS K\pi}_D\kappa^{\KS K\pi}_D\cos(\delta^{D\pi}_B-\delta^{\KS K\pi}_D)\cos\phi_3}{(r^{D\pi}_B)^2+(r^{\KS K\pi}_D)^2+2r^{D\pi}_Br^{\KS K\pi}_D\kappa^{\KS K\pi}_D\cos(\delta^{D\pi}_B+\delta^{\KS K\pi}_D)\cos\phi_3}.
\end{split}
\end{align}
\end{itemize}

\notocsection{Input observable values, uncertainties, and uncertainties correlations}
\label{app:inputs}
We list the input observables' values, uncertainties, as well as correlations between uncertainties.

\notocsubsection{$B^+\to Dh^+, D\to\KS h^-h^+$ analysis}
The values and uncertainties are taken from Ref.~\cite{niharika}. These are

\begin{equation} \label{eq:xy_param}
    \begin{split}
        x_{-}^{DK} &=  \phantom{-}0.0924 \pm 0.0327 \pm 0.0029, \\
        y_{-}^{DK} &=  \phantom{-}0.1000 \pm 0.0420 \pm 0.0074, \\
        x_{+}^{DK} &= -0.1128 \pm 0.0315 \pm 0.0029, \\
        y_{+}^{DK} &= -0.0455 \pm 0.0420 \pm 0.0055, \\
        x_{\xi}^{D\pi} &= -0.1109 \pm 0.0475 \pm 0.0085, \\
        y_{\xi}^{D\pi} &= -0.0790 \pm 0.0544 \pm 0.0083, \\
    \end{split}
\end{equation}
where the first uncertainty is statistical and the second one includes systematic uncertainty and the additional uncertainty from the strong-interaction phase difference in $D\to\KS\pip\pim$ decays. The statistical and systematic correlation matrices are given in Tables~\ref{tab:KshhStatCorr} and \ref{tab:KshhSystCorr}. 

\begin{table}[!h]
    \centering
    \caption{Correlation matrix of the statistical uncertainties for the $B^+ \to Dh^+, D \to \KS h^-h^+$ observables.}
    \vspace{0.5cm}
    \begin{tabular}{c | c c c c c c }
    \hline
      & $x_{-}^{DK}$ & $y_{-}^{DK}$ & $x_{+}^{DK}$ & $y_{+}^{DK}$ & $x_{\xi}^{D\pi}$ & $y_{\xi}^{D\pi}$\\
    \hline
     $x_{-}^{DK}$ &     1.000   & $-0.204$&  $-0.051$& $\phantom{-}$0.063& $\phantom{-}$0.365&  $-0.151$ \\
     $y_{-}^{DK}$ & &  1.000   & $\phantom{-}$0.014&  $-0.051$& $-0.090$& $\phantom{-}$0.404 \\
     $x_{+}^{DK}$ & & &  $\phantom{-}$1.000   &  $\phantom{-}$0.152&  $-0.330$& $-0.057$ \\
     $y_{+}^{DK}$ & & & &  $\phantom{-}$1.000&    $\phantom{-}$0.026&  $-0.391$ \\
     $x_{\xi}^{D\pi}$ & & & & &  $\phantom{-}$1.000&    $\phantom{-}$ 0.080\\
     $y_{\xi}^{D\pi}$ & & & & & & $\phantom{-}$1.000 \\
    \hline
    \end{tabular}
    \label{tab:KshhStatCorr}
\end{table}

\begin{table}[!h]
    \centering
    \caption{Correlation matrix of the systematic uncertainties for the $B^+ \to Dh^+, D \to \KS h^-h^+$ observables.}
    \vspace{0.5cm}
    \begin{tabular}{c | c c c c c c }
    \hline
      & $x_{-}^{DK}$ & $y_{-}^{DK}$ & $x_{+}^{DK}$ & $y_{+}^{DK}$ & $x_{\xi}^{D\pi}$ & $y_{\xi}^{D\pi}$\\
    \hline
      $x_{-}^{DK}$ &1.000 & 0.104 & 0.228 & $\phantom{-}$0.335 & $\phantom{-}$0.248 & $\phantom{-}$0.145 \\
      $y_{-}^{DK}$ & & 1.000 & 0.199 & $-0.119$ & $-0.410$ & $-0.103$ \\ 
      $x_{+}^{DK}$ & & & 1.000 & $\phantom{-}$0.423 & $\phantom{-}$0.063 & $-0.375$ \\ 
      $y_{+}^{DK}$ & & & & $\phantom{-}$1.000 & $\phantom{-}$0.173 & $-0.089$ \\ 
      $x_{\xi}^{D\pi}$ & & & & & $\phantom{-}$1.000 & $\phantom{-}$0.566 \\ 
      $y_{\xi}^{D\pi}$ & & & & & & $\phantom{-}$1.000 \\
    \hline
    \end{tabular}
    \label{tab:KshhSystCorr}
\end{table}

\notocsubsection{$B^+\to Dh^+, D\to\KS\pi^-\pi^+\piz$ analysis}
The values and uncertainties are taken from Ref.~\cite{resmi1}. Those are

\begin{equation} \label{eq:xy_param2}
\begin{split}
        x_{-}^{DK} &=  \phantom{-}0.095 \pm 0.121 \pm 0.029, \\
        y_{-}^{DK} &=  \phantom{-}0.354 \pm 0.170 \pm 0.045, \\
        x_{+}^{DK} &= -0.030 \pm 0.121 \pm 0.026, \\
        y_{+}^{DK} &=  \phantom{-}0.220 \pm 0.376 \pm 0.079, \\
        x_{-}^{D\pi} &= -0.014 \pm 0.021 \pm 0.021, \\
        y_{-}^{D\pi} &= -0.033 \pm 0.059 \pm 0.023, \\
        x_{+}^{D\pi} &=  \phantom{-}0.039 \pm 0.024 \pm 0.020, \\
        y_{+}^{D\pi} &= -0.196 \pm 0.069 \pm 0.048, \\
    \end{split}
\end{equation}
where the first uncertainty is statistical and the second one includes systematic uncertainty and the additional uncertainty from the strong-interaction phase difference in $D\to\KS\pip\pim\piz$ decays.
The statistical correlation matrix is given in Table~\ref{tab:KspipipizStatCorr}. The systematic correlation matrix was not reported for this measurement. 

\begin{table}[!h]
    \centering
    \caption{Correlation matrix of the statistical uncertainties for the $B^+ \to Dh^+, D \to \KS \pi^-\pi^+\pi^0$ observables.}
    \vspace{0.5cm}
    \begin{tabular}{c | c c c c c c c c}
    \hline
      & $x_{-}^{DK}$ & $y_{-}^{DK}$ & $x_{+}^{DK}$ & $y_{+}^{DK}$ & $x_{-}^{D\pi}$ & $y_{-}^{D\pi}$ & $x_{+}^{D\pi}$ & $y_{+}^{D\pi}$\\
    \hline
     $x_{-}^{DK}$ &  1.000   & 0.486& 0.172& $-0.231$& 0.000 &0.000 &0.000 &0.000 \\ 
     $y_{-}^{DK}$ &  &  1.000   &$-0.127$& 0.179& 0.000 &0.000 &0.000 &0.000 \\
     $x_{+}^{DK}$ & & &  1.000   &  0.365& 0.000 &0.000 &0.000 &0.000 \\
     $y_{+}^{DK}$ & & & &  1.000 &  0.000 &0.000 &0.000 &0.000 \\
     $x_{-}^{D\pi}$& & & & & 1.000 & $-0.364$& 0.314& 0.050\\
     $y_{-}^{D\pi}$ & & & & & & 1.000 & 0.347& 0.055\\
     $x_{+}^{D\pi}$ &  & & & & & & 1.000 & $-0.032$\\
     $y_{+}^{D\pi}$& & & & & & & & 1.000\\
    \hline
    \end{tabular}
    \label{tab:KspipipizStatCorr}
\end{table}

\notocsubsection{$B^+\to D^*K^+, D\to\KS\pim\pip$ analysis}
The values and uncertainties are taken from Ref.~\cite{anton}. Those are
\begin{equation} \label{eq:xy_param3}
    \begin{split}
        x_{-}^{(D\piz) K} &= \phantom{-}0.024 \pm 0.140 \pm 0.018, \\
        y_{-}^{(D\piz) K} &= -0.243 \pm 0.137 \pm 0.022, \\
        x_{+}^{(D\piz) K} &= \phantom{-}0.133 \pm 0.083 \pm 0.018, \\
        y_{+}^{(D\piz) K} &= \phantom{-}0.130 \pm 0.120 \pm 0.022, \\
        x_{-}^{(D\gamma) K} &= \phantom{-}0.144 \pm 0.208 \pm 0.025, \\
        y_{-}^{(D\gamma) K} &= \phantom{-}0.196 \pm 0.215 \pm 0.037, \\
        x_{+}^{(D\gamma) K} &= -0.006 \pm 0.147 \pm 0.025, \\
        y_{+}^{(D\gamma) K} &= -0.190 \pm 0.177 \pm 0.037, \\
    \end{split}
\end{equation}
where the first uncertainty is statistical and the second one systematic. The uncertainty from the $D\to\KS\pip\pim$ decay model is unknown. The statistical correlation matrix is given in Table~\ref{tab:DstKKspipiCorr}. The systematic correlation matrix was not reported for this measurement. 

\begin{table}[!h]
    \centering
    \caption{Correlation matrix of the statistical uncertainties for the $B^+ \to \Dstar K^+, \Dstar\to D\piz/\gamma, D\to \KS\pim\pip$ observables.}
    \vspace{0.5cm}
    \begin{tabular}{c |c c c c c c c c}
    \hline
      & $x_{-}^{(D\piz) K}$  & $y_{-}^{(D\piz) K}$ & $x_{+}^{(D\piz) K}$ & $y_{+}^{(D\piz) K}$ & $x_{-}^{(D\gamma) K}$ & $y_{-}^{(D\gamma) K}$ & $x_{+}^{(D\gamma) K}$ & $y_{+}^{(D\gamma) K}$\\
    \hline
      $x_{-}^{(D\piz) K}$ & 1.000   & 0.440& 0.000 &0.000 &0.000 &0.000 &0.000 &0.000 \\
      $y_{-}^{(D\piz) K}$ & & 1.000 &     0.000 &0.000 &0.000 &0.000 &0.000 &0.000 \\
      $x_{+}^{(D\piz) K}$ & & & 1.000   &$-0.101$& 0.000 &0.000 &0.000 &0.000 \\
      $y_{+}^{(D\piz) K}$ &  & & & 1.000 &      0.000 &0.000 &0.000 &0.000 \\
      $x_{-}^{(D\gamma) K}$ &  & & & &  1.000   &$-0.207$& 0.000 &0.000 \\
      $y_{-}^{(D\gamma) K}$ &  & & & & & 1.000 &     0.000 &0.000 \\
      $x_{+}^{(D\gamma) K}$ &  & & & & & &1.000   & 0.080\\
      $y_{+}^{(D\gamma) K}$ &  & & & & & & & 1.000     \\
    \hline
    \end{tabular}
    \label{tab:DstKKspipiCorr}
\end{table}    

\notocsubsection{$B^+\to Dh^+, D\to K^+\pim$ analysis}
The values and uncertainties are taken from Ref.~\cite{horii}. Those are
\begin{equation} \label{eq:ads_param1}
\begin{split}
        R_{K\pi}^{DK} &= \phantom{-}0.0163 \pm 0.0042 \pm 0.0010, \\
        A_{K\pi}^{DK} &= -0.39 \pm 0.27 \pm 0.04, \\
        R_{K\pi}^{D\pi} &= \phantom{-}0.00328 \pm 0.00037 \pm 0.00015, \\
        A_{K\pi}^{D\pi} &= -0.04 \pm 0.11 \pm 0.02, \\
    \end{split}
\end{equation}
where the first uncertainty is statistical and the second one systematic.
The statistical correlation matrix is given in Table~\ref{tab:ADSKpiCorr}. The systematic correlation matrix was not reported for this measurement. 

\begin{table}[!h]
    \centering
    \caption{Correlation matrix of the statistical uncertainties for the $B^+\to Dh^+, D\to K^+\pim$ observables.}
    \vspace{0.5cm}
    \begin{tabular}{c |c c c c }
    \hline
      & $R_{K\pi}^{DK}$  & $A_{K\pi}^{DK}$ & $R_{K\pi}^{D\pi}$ & $A_{K\pi}^{D\pi}$ \\
    \hline
    $R_{K\pi}^{DK}$ & 1.000 & 0.242 & 0.000 & 0.000 \\
    $A_{K\pi}^{DK}$ &  & 1.000 & 0.000 & 0.000 \\
    $R_{K\pi}^{D\pi}$ &  & & 1.000 & $-0.032$ \\
    $A_{K\pi}^{D\pi}$ &  &  & & 1.000\\
    \hline
    \end{tabular}
    \label{tab:ADSKpiCorr}
\end{table}    

\notocsubsection{$B^+\to Dh^+, D\to K^+\pim\piz$ analysis}
The values and uncertainties are taken from Ref.~\cite{nayak}. Those are
\begin{equation} \label{eq:ads_param2}
\begin{split}
        R_{K\pi\pi^0}^{DK} &= (1.98 \pm 0.62 \pm 0.24) \times 10^{-2}, \\
        A_{K\pi\pi^0}^{DK} &= \phantom{(}0.41 \pm 0.307 \pm 0.05, \\
        R_{K\pi\pi^0}^{D\pi} &= (0.19 \pm 0.05 \pm 0.02) \times 10^{-2}, \\
        A_{K\pi\pi^0}^{D\pi} &= \phantom{(}0.16 \pm 0.27 \pm 0.04, \\
    \end{split}
\end{equation}
where the first uncertainty is statistical and the second one systematic.
The statistical and systematic correlation matrices were not reported for this measurement. 

\notocsubsection{$B^+\to Dh^+, D\to \KS\piz, K^-K^+$ analysis}
The values and uncertainties are taken from Ref.~\cite{yi-analysis}. Those are
\begin{equation} \label{eq:glw_param}
\begin{split}
        A_{\CP-}^{DK} &= -0.167 \pm 0.057 \pm 0.006, \\
        R_{\CP-}^{DK} &= \phantom{-}1.151 \pm 0.074 \pm 0.019, \\
        A_{\CP+}^{DK} &= \phantom{-}0.125 \pm 0.058 \pm 0.014, \\
        R_{\CP+}^{DK} &= \phantom{-}1.164 \pm 0.081 \pm 0.036, \\
    \end{split}
\end{equation}
where the first uncertainty is statistical and the second one systematic.
The statistical and systematic correlation matrices are given in Tables~\ref{tab:GLWStatCorr} and \ref{tab:GLWSystCorr}. 

\begin{table}[!h]
    \centering
    \caption{Correlation matrix of the statistical uncertainties for the $B^+ \to DK^+, D \to \KS \pi^0, K^-K^+$ observables.}
    \vspace{0.5cm}
    \begin{tabular}{c| c c c c  }
    \hline
      & $A_{\CP-}^{DK}$ & $R_{\CP-}^{DK}$ & $A_{\CP+}^{DK}$ & $R_{\CP+}^{DK}$ \\
    \hline
     $A_{\CP-}^{DK}$ &  1.000   & 0.056&  0.000& 0.000\\
     $R_{\CP-}^{DK}$ &  &  1.000   & 0.000&  $-0.081$\\
     $A_{\CP+}^{DK}$ &  & &  1.000   &  0.060\\
     $R_{\CP+}^{DK}$ & &  &  &  1.000 \\
    \hline
    \end{tabular}
    \label{tab:GLWStatCorr}
\end{table}
  
\begin{table}[!h]
    \centering
    \caption{Correlation matrix of the systematic uncertainties for the $B^+ \to DK^+, D \to \KS \pi^0, K^-K^+$ observables.}
    \vspace{0.5cm}
    \begin{tabular}{c |c c c c  }
    \hline
      & $A_{\CP-}^{DK}$ & $R_{\CP-}^{DK}$ & $A_{\CP+}^{DK}$ & $R_{\CP+}^{DK}$ \\
    \hline
     $A_{\CP-}^{DK}$ &  1.000   &  $-0.490$& 0.540&  0.005  \\
     $R_{\CP-}^{DK}$ &  &  1.000   &  $-0.128$& $-0.063$ \\
     $A_{\CP+}^{DK}$ &  &  &  1.000   & 0.342 \\ 
     $R_{\CP+}^{DK}$ &  & &  &  1.000 \\
    \hline
    \end{tabular}
    \label{tab:GLWSystCorr}
\end{table}

\notocsubsection{$B^+\to D^*h^+, D\to \KS\piz, K^-K^+, \KS\phi, \KS\omega, \pim\pip$ analysis}
The values and uncertainties are taken from Ref.~\cite{glw2}. Those are
\begin{equation} \label{eq:glw_param2}
\begin{split}
        A_{\CP-}^{DK} &= \phantom{-}0.13 \pm 0.30 \pm 0.08, \\
        R_{\CP-}^{DK} &= \phantom{-}1.15 \pm 0.31 \pm 0.12, \\
        A_{\CP+}^{DK} &= -0.20 \pm 0.22 \pm 0.04, \\
        R_{\CP+}^{DK} &= \phantom{-}1.41 \pm 0.25 \pm 0.06, \\
    \end{split}
\end{equation}
where the first uncertainty is statistical and the second one systematic.
The statistical and systematic correlation matrices were not reported for this measurement. 

\notocsubsection{$B^+\to Dh^+, D\to \KS K^-\pip$ analysis}
The values and uncertainties are taken from Ref.~\cite{gls}. Those are
\begin{equation} \label{eq:gls_kstark}
\begin{split}
  A_{\text{SS}}^{{\it D K}}           &= 0.055 \pm 0.119 \pm 0.020,\\
  A_{\text{OS}}^{{\it D K}}           &= 0.231 \pm 0.184 \pm 0.014,\\
  A_{\text{SS}}^{{\it D \pi}}           &= 0.046 \pm 0.029 \pm 0.016,\\
  A_{\text{OS}}^{{\it D \pi}}           &= 0.009 \pm 0.046 \pm 0.009,\\
  R_{\text{SS}}^{{\it D K}/{\it D \pi}}    &= 0.093 \pm 0.012 \pm 0.005,\\
  R_{\text{OS}}^{{\it D K}/{\it D \pi}}    &= 0.103 \pm 0.020 \pm 0.006,\\
  R_{\text{SS}/\text{OS}}^{{\it D \pi}} &= 2.412 \pm 0.132 \pm 0.019,
    \end{split}
\end{equation}
where the first uncertainty is statistical and the second one systematic.
The statistical and systematic correlation matrices are given in Tables~\ref{tab:GLSStatCorr} and \ref{tab:GLSSystCorr}. 

\begin{table}[!h]
    \centering
    \caption{Correlation matrix of the statistical uncertainties for the $B^+ \to Dh^+, D \to \KS K^-\pip$ observables.}
    \vspace{0.5cm}
    \begin{tabular}{c |c c c c c c c }
    \hline
      & $A_{\text{SS}}^{{\it D K}}$ & $A_{\text{OS}}^{{\it D K}}$ & $A_{\text{SS}}^{{\it D \pi}}$ & $A_{\text{OS}}^{{\it D \pi}}$ & $R_{\text{SS}}^{{\it D K}/{\it D \pi}}$ & $R_{\text{OS}}^{{\it D K}/{\it D \pi}}$ & $R_{\text{SS}/\text{OS}}^{{\it D \pi}}$ \\
    \hline
     $A_{\text{SS}}^{{\it D K}}$ &  1.000 & 0.003 & $-0.012$ & 0.001 & $-0.052$ & $-0.013$ & 0.002 \\ 
     $A_{\text{OS}}^{{\it D K}}$ &   & 1.000 & 0.001 & $-0.011$ & $-0.004$ & $-0.034$ & 0.002 \\ 
     $A_{\text{SS}}^{{\it D \pi}}$ &  &  & 1.000 & 0.001 & 0.002 & $-0.004$ & $-0.011$ \\ 
     $A_{\text{OS}}^{{\it D \pi}}$ &  &  &  & 1.000 & $-0.002$ & $-0.002$ & 0.014 \\
    $R_{\text{SS}}^{{\it D K}/{\it D \pi}}$ &  &  &  &  & 1.000 & 0.034 & $-0.133$ \\ 
     $R_{\text{OS}}^{{\it D K}/{\it D \pi}}$ &  &  &  &  &  & 1.000 & 0.208 \\
     $R_{\text{SS}/\text{OS}}^{{\it D \pi}}$&  &  &  &  &  &  & 1.000 \\
     
    \hline
    \end{tabular}
    \label{tab:GLSStatCorr}
\end{table}

\begin{table}[!h]
    \centering
    \caption{Correlation matrix of the statistical uncertainties for the $B^+ \to Dh^+, D \to \KS K^-\pip$ observables.}
    \vspace{0.5cm}
    \begin{tabular}{c |c c c c c c c }
    \hline
      & $A_{\text{SS}}^{{\it D K}}$ & $A_{\text{OS}}^{{\it D K}}$ & $A_{\text{SS}}^{{\it D \pi}}$ & $A_{\text{OS}}^{{\it D \pi}}$ & $R_{\text{SS}}^{{\it D K}/{\it D \pi}}$ & $R_{\text{OS}}^{{\it D K}/{\it D \pi}}$ & $R_{\text{SS}/\text{OS}}^{{\it D \pi}}$ \\
    \hline
      $A_{\text{SS}}^{{\it D K}}$ & 1.000 & 0.195 & 0.047 & 0.013 & 0.120 & $-0.053$ & 0.192 \\
      $A_{\text{OS}}^{{\it D K}}$ &  & 1.000 & 0.038 & 0.004 & 0.344 & 0.210 & 0.007 \\ 
      $A_{\text{SS}}^{{\it D \pi}}$&  &  & 1.000 & 0.024 & $-0.004$ & $-0.037$ & 0.018 \\
      $A_{\text{OS}}^{{\it D \pi}}$ &  &  &  & 1.000 & $-0.017$ & $-0.024$ & 0.006 \\
      $R_{\text{SS}}^{{\it D K}/{\it D \pi}}$ &  &  &  &  & 1.000 & 0.915 & 0.015 \\
      $R_{\text{OS}}^{{\it D K}/{\it D \pi}}$ &  &  &  &  &  & 1.000 & $-0.097$ \\ 
      $R_{\text{SS}/\text{OS}}^{{\it D \pi}}$&  &  &  &  &  &  & 1.000 \\
    \hline
    \end{tabular}
    \label{tab:GLSSystCorr}
\end{table}

\notocsection{External inputs' values, uncertainties, and uncertainties correlations}
\label{app:charminputs}
\notocsubsection{Inputs from global fit to $\Dz-\Dzb$ mixing data}
The values and uncertainties are taken from Ref.~\cite{hflav}. Those are
\begin{equation} \label{eq:hflav_param}
\begin{split}
        x_D &= (0.407 \pm 0.044)\%, \\
        y_D &= (0.647 \pm 0.024)\%, \\
        \delta_D^{K\pi} &= (191.7 \pm 3.7)^\circ, \\
        (r_D^{K\pi})^2 &= (3.44 \pm 0.02)\times 10^{-3}, 
    \end{split}
\end{equation}
where the uncertainty includes both statistical and systematic uncertainties. 
The correlation matrix is given in Table~\ref{tab:DKpiCorr}.
\begin{table}[!h]
    \centering
    \caption{Correlation matrix for all uncertainties of the input variables from global fit to $\Dz-\Dzb$ mixing data.}
    \vspace{0.5cm}
    \begin{tabular}{c |c c c c}
    \hline
      & $x_D$ & $y_D$ & $\delta_D^{K\pi}$ & $(r_D^{K\pi})^2$\\
    \hline
     $x_D$ & 1.000 & -0.030 & -0.049 & 0.023 \\
     $x_D$ & & 1.000 & 0.867 & 0.145\\
     $\delta_D^{K\pi}$& & & 1.000 & 0.534 \\
     $(r_D^{K\pi})^2$& & & & 1.000\\
    \hline
    \end{tabular}
    \label{tab:DKpiCorr}
\end{table}

\notocsubsection{Input for $D \to K^+\pim$}
The values and uncertainties are taken from Ref.~\cite{CharmKpiBESIII}. Those are
\begin{equation} \label{eq:kpi_param}
\begin{split}
        r_D^{K\pi}\cos(\delta_D^{K\pi}) &= -0.0562 \pm 0.0081 \pm 0.0051, \\
        r_D^{K\pi}\sin(\delta_D^{K\pi}) &= -0.011 \pm 0.012 \pm 0.0076, 
    \end{split}
\end{equation}
where the first uncertainty is statistical and the second one systematic.
The correlation between these two quantities is 0.02.

\notocsubsection{Input for $D \to K^+\pim\piz$}
The values and uncertainties are taken from ~\cite{CharmKpipi0BESIII}.The values used are
\begin{equation} \label{eq:kpipi0_param1}
\begin{split}
        r_D^{K\pi\piz} &= 0.0441 \pm 0.0011 \\
        \kappa_D^{K\pi\piz} &= 0.79 \pm 0.04, \\
        \delta_D^{K\pi\piz} &= (196 \pm 11)^\circ,
    \end{split}
\end{equation}
where the uncertainty includes both statistical and systematic uncertainties. 
The correlation matrix is given in Table~\ref{tab:DKpipi0Corr2}.

\begin{table}[!h]
    \centering
    \caption{Correlation matrix for all uncertainties of the $D \to K^+\pim\piz$ channel parameters.}
    \vspace{0.5cm}
    \begin{tabular}{c| c c c}
    \hline
      & $R_D^{K\pi\piz}$ & $\delta_D^{K\pi\piz}$ & $r_D^{K\pi\piz}$ \\
    \hline
      $R_D^{K\pi\piz}$ & 1.00 & 0.19 & $-0.01$ \\
      $\delta_D^{K\pi\piz}$ & & 1.00 & $0.25$ \\
      $r_D^{K\pi\piz}$ &  &  & 1.00 \\
    \hline
    \end{tabular}
    \label{tab:DKpipi0Corr2}
\end{table}

\notocsubsection{Input for $D \to \KS K^-\pip$}
The values and uncertainties are taken from Ref.~\cite{glsCharmCLEO}. Those are
\begin{equation} \label{eq:kskpi_param1}
\begin{split}
        (r_D^{\KS K\pi})^2 &= 0.356 \pm 0.034 \pm 0.007, \\
        \kappa_D^{\KS K\pi} &= 0.94 \pm 0.12, \\
        \delta_D^{\KS K\pi} &= (-16.6 \pm 18.4)^\circ.
    \end{split}
\end{equation}
For $(r_D^{\KS K\pi})^2$, the first uncertainty is statistical, and the second one is systematic.
For  $\kappa_D^{\KS K\pi}$ and $\delta_D^{\KS K\pi}$, the uncertainty includes both statistical and systematic uncertainties. 
The correlation matrix is given in Table~\ref{tab:DKsKpiCorr}.

\begin{table}[!h]
    \centering
    \caption{Correlation matrix for all uncertainties of the $D \to \KS K^-\pip$ channel parameters from CLEO.}
    \vspace{0.5cm}
    \begin{tabular}{c| c c c}
    \hline
      & $(r_D^{\KS K\pi})^2$ & $\delta_D^{\KS K\pi}$ & $\kappa_D^{\KS K\pi}$ \\
    \hline
      $(r_D^{\KS K\pi})^2$ & 1.0 & 0.0 & 0.0 \\
      $\delta_D^{\KS K\pi}$ &  & 1.0 & $-0.6$ \\
      $\kappa_D^{\KS K\pi}$ &  &  & 1.0 \\
    \hline
    \end{tabular}
    \label{tab:DKsKpiCorr}
\end{table}

In addition, the following input from Ref.~\cite{glsCharmLHCb} is used,
\begin{equation} \label{eq:kskpi_param2}
\begin{split}
        (r_D^{\KS K\pi})^2 &= 0.370 \pm 0.003 \pm 0.012, 
    \end{split}
\end{equation}
where the first uncertainty is statistical and the second one systematic.
The value 
\begin{equation} \label{eq:kskpi_param3}
\begin{split}
        R(\BR(B^- \to D^0K^-)/\BR(B^- \to D^0\pi^-)) &= 0.0789 \pm 0.0027 
    \end{split}
\end{equation}
from the PDG~\cite{pdg} is also used, where the uncertainty includes both statistical and systematic uncertainties. 

\end{document}